\documentclass[aps,prl,reprint,superscriptaddress,longbibliography,floatfix]{revtex4-2}
\usepackage{amsmath,amssymb,graphicx,bm}
\usepackage{mathrsfs}
\usepackage[hidelinks]{hyperref}
\newcommand{\tr}{\operatorname{Tr}}

\newcommand{\kc}{k_{\mathrm c}}
\begin{document}
\title{How Many Autonomous Encoded Qubits Force an Isotropic Noise Response?}
\author{Xu Chen}
\email{cnc@hebust.edu.cn}
\affiliation{School of Sciences, Hebei University of Science and Technology, Shijiazhuang 050018, Hebei, People's Republic of China}

\author{Xue Ma}
\affiliation{Network Management Center,
China Mobile Communications Group Hebei Co., Ltd., Shijiazhuang 050000, Hebei, People's Republic of China}
\begin{abstract}
Depolarizing noise scales trace distances by a common factor. An encoded qubit is autonomous if its reduced input state determines its output. For $D=2^n\ge4$, autonomy forces this isotropy at a sharp threshold of $(D^2+5)/3$ generic encodings, two more than state tomography requires. For Pauli noise, one generic encoding suffices. With one-bit readouts and no quantum memory or reference, switching encodings cuts the worst case cost for the Pauli test by a factor of $D$, up to logarithms. For any family, a channel with leakage $O(D^{-3/2})$ can transmit one bit perfectly and erase another.
\end{abstract}
\maketitle
\raggedbottom
\clubpenalty=10000
\widowpenalty=10000
Global depolarization attenuates the distinguishability of every pair of states by the same factor. This property underlies noise mitigation by rescaling expectation values~\cite{Urbanek2021,Vovrosh2021} and white noise models of random circuit experiments~\cite{Arute2019,Dalzell2024}. A stochastic phase flip illustrates what can go wrong. The two $Z$ basis states remain perfectly distinguishable, while the two $X$ basis states become harder to distinguish. Which measurements on encoded qubits can certify that the noise acting on the register has no preferred basis? 

\begin{figure}[tb]
 \centering\includegraphics[width=\columnwidth]{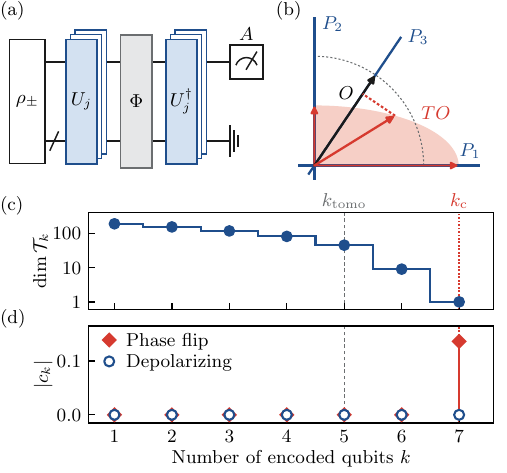}
 \caption{Autonomy and the isotropy threshold. (a) The stacked gates represent $k$ overlapping encodings. Qubit $j$ is
autonomous when any two inputs $\rho_\pm$ with the same logical marginal give the same expectation for every logical Pauli $A$ after decoding. (b) Planar analog of Eq.~\eqref{eq:autonomy}. $T$ preserves both coordinate axes and maps the unit quarter disk to the shaded region. The red dotted segment shows the leakage $(I-P_3)TO$ from the oblique line. Only $T=\lambda I$ preserves all three lines. (c) For one sequence of Haar encoders at $D=4$, $\dim\mathcal T_k$ is $45$ at the tomographic count ($k=5$), $9$ at $k=6$, and $1$ at $k=\kc=7$, where only the common rate $\lambda$ remains. (d) Largest half difference $|c_k|$ between the two mean readouts in (a), over three logical axes and twelve complementary Pauli probes. The first 6 encodings span all 15 axes and give zero for both channels. The seventh gives $|c_7|\simeq0.1364$ for the $Z\otimes I$ flip ($p=1/4$, diamonds) and zero for depolarization ($\lambda=11/15$, circles). }  \label{fig:messages}
\end{figure}

An encoded qubit is specified by its observable algebra~\cite{Zanardi2001,Zanardi2004}. We call the qubit autonomous if its output depends only on its input, even when the qubit and its spectators are correlated. In this case, the spectators cannot signal to the qubit~\cite{Beckman2001,Eggeling2002}, and the adjoint channel preserves its observable algebra~\cite{HasenoehrlCaro2023}. Spectator dependence is a signature of crosstalk~\cite{Rudinger2019,Sarovar2020}. Noiseless subsystems~\cite{KnillLaflammeViola2000,Kribs2005} and qubits that decay independently of their spectators are both autonomous.

Different choices of tensor factors define different encoded qubits in the same register~\cite{Zanardi2001,Zanardi2004}. Dimension tests for overlapping qubits assume that Pauli observables of different qubits approximately commute~\cite{Chao2017}. The encoded qubits need not commute, even approximately, and every autonomy test constrains the same noise channel. For a large enough family, these conditions force the channel to scale every state difference by the same factor.

\begin{figure}[!tb]
 \centering\includegraphics[width=\columnwidth]{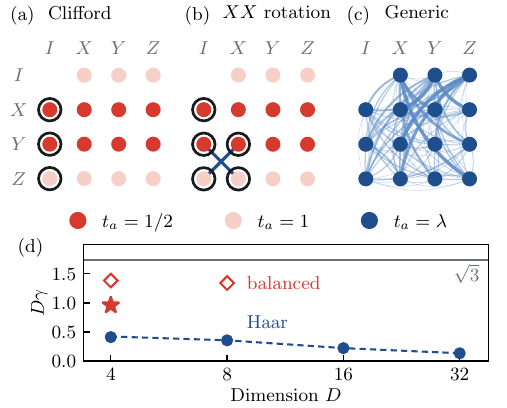}
 \caption{How one autonomous encoded qubit constrains Pauli decay rates. (a)--(c) Nodes are the 15 nonidentity Pauli axes, arranged by the first and second tensor factors. Fill denotes the phase flip rates of Fig.~\ref{fig:messages} or a common rate $\lambda$. Outlines mark support on the logical axes.  Autonomy requires $t_a=t_b$ along each edge. (a) A Clifford encoder gives no edges. (b) The rotation $e^{-i\pi XX/8}$ detects the $Z\otimes I$ flip because the rates differ across both edges. It misses an $X\otimes I$ flip, whose rates agree along both edges. (c) The rational encoder~\cite{SupplementalMaterial} (qubits exchanged) links all 105 pairs, forcing a common rate $\lambda$. Line weight increases with $|P_{ab}|^2$. (d) Circles and dashes show the medians of $D\gamma$ and $D\gamma_w$ over 512 Haar encoders per dimension. The upper limit is $\sqrt3$. The star denotes the certified bound $4\gamma>0.96$ for (c), and diamonds denote balanced encoders. }
 \label{fig:pauli}
\end{figure}

\begin{figure}[tb]
 \centering\includegraphics[width=\columnwidth]{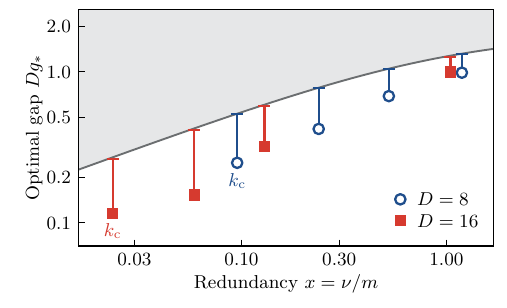}
\caption{Optimal gap $Dg_*$ versus fractional redundancy
$x=\nu/m$. Circles ($D=8$) and squares ($D=16$) mark certified lower bounds. The upper bounds (caps) from Eq.~\eqref{eq:relation_bound} exclude the shaded region for both dimensions. At the threshold $k=\kc$, $x=6/m$. }
 \label{fig:precision}
\end{figure}

We answer this question for dimension $D=2^n$ with $n\ge2$. For arbitrary noise, the sharp threshold is $(D^2+5)/3$ generic encoded qubits, exactly two more than state tomography requires. With one encoding fewer, anisotropic noise can pass every autonomy test of a tomographically complete family [Fig.~\ref{fig:messages}(d)]. Pauli noise needs only one suitably rotated qubit (Fig.~\ref{fig:pauli}). Robust certification asks for more. Additional encodings reduce the amplification of leakage from order $D^2$ at threshold to $D$ at fixed fractional redundancy (Fig.~\ref{fig:precision}), yet a channel with leakage of $D^{-3/2}$ can still transmit a bit perfectly in one basis and erase it in another. For Pauli noise, switching the encoding between trials saves a factor of $D$ in channel uses, up to logarithms.

\emph{Autonomy and leakage.}---
A completely positive trace-preserving map $\Phi$ describes the device's evolution from preparation to a fixed observation time. An encoder $U$ selects the encoded qubit $U(M_2\otimes I_{D/2})U^\dagger$. We expand the adjoint channel in the Pauli basis of $m=D^2-1$ traceless observables, orthonormal for $\operatorname{Tr}(AB)/D$. The response $T$ is the block indexed by the traceless Pauli operators. Let $P_j$ project onto the three logical Pauli operators of qubit $j$. The qubit is autonomous exactly when 
\begin{equation}
 (I-P_j)\,T P_j=0 .
 \label{eq:autonomy}
\end{equation}
Prepare two states by shifting $I_D/D$ by equal amounts in opposite directions orthogonal to the logical algebra. Their logical inputs agree. The difference between their logical Pauli readouts, divided by the preparation separation, measures one entry of $(I-P_j)TP_j$ [Figs.~\ref{fig:messages}(a) and~\ref{fig:messages}(b)]~\cite{SupplementalMaterial}. For $k$ encoded qubits, define the root-mean-square leakage $e$ and the gap $g$ by
\begin{equation}
 e(T)^2=\frac1k\sum_{j=1}^k\|(I-P_j)TP_j\|_F^2,\qquad
 g=\min_{\substack{\operatorname{Tr}X=0\\ \|X\|_F=1}}e(X),
 \label{eq:leak}
\end{equation}
where $X$ ranges over real $m\times m$ matrices. The gap is positive exactly when autonomy of the whole family forces a scalar response. Let $\lambda=\operatorname{Tr}T/m$ be the mean rate. Since $T-\lambda I_m$ is traceless and has the same leakage as $T$, for $g>0$ the anisotropy satisfies
\begin{equation}
 R=\|T-\lambda I_m\|_F\le\frac{e(T)}{g},
 \label{eq:master}
\end{equation} 
Unitarity benchmarking gives $u=\|T\|_F^2/m=\lambda^2+R^2/m$~\cite{WallmanUnitarity2015,Dirkse2019}. Together with $\lambda$, it fixes the magnitude $R$. For $\lambda\ne0$, dividing output state differences by $\lambda$ gives a relative Hilbert-Schmidt error of at most $R/|\lambda|$.

\emph{Sharp threshold.}---
A scalar response $T=\lambda I_m$ acts on state differences as
\begin{equation}
 \Phi(\rho)-\Phi(\sigma)=\lambda(\rho-\sigma),
 \label{eq:commonrate}
\end{equation}
 for all states $\rho,\sigma$. Every trace distance scales by $|\lambda|$. For two equiprobable inputs, the optimal probability of guessing the input from the output is therefore $1/2+|\lambda|\,\|\rho-\sigma\|_1/4$~\cite{Helstrom1967,Watrous2018}. For $\lambda<1$, $\Phi(\rho)=\lambda\rho+(1-\lambda)\tau$, where $\tau$ is its unique fixed state. If $\Phi$ is unital, it fixes $I_D/D$ and hence equals the depolarizing channel $\mathcal D_\lambda$.

\emph{Theorem 1.}---Let $D=2^n\ge4$. The sharp generic threshold for certifying an isotropic response from encoded qubits autonomy is
\begin{equation}
 \kc=\frac{D^2+5}{3}.
 \label{eq:threshold}
\end{equation}
For every family of $k<\kc$ encoded qubits, there exists a channel makes all $k$ autonomous and scales the trace distances of two state pairs by different factors. For Haar-almost every family of $k\ge\kc$ encoders, autonomy of all $k$ qubits forces Eq.~\eqref{eq:commonrate}. Removing any one encoded qubit from a forcing family leaves a tomographically complete family.

The threshold follows from relations among the logical axes. For a spanning family, the $3k$ axes obey $\nu=3k-m$ independent linear relations. State tomography from encoded qubits marginals, studied for two qubits by Petz \emph{et al.}~\cite{PRPetz}, requires $k\ge k_{\rm tomo}=m/3$, and generic encodings attain this count~\cite{SupplementalMaterial}. Spanning is not enough: in the planar analog of Fig.~\ref{fig:messages}(b), two invariant lines span the plane yet allow any diagonal response, and a third line forces $T=\lambda I$. Equation~\eqref{eq:autonomy} gives $3(m-3)$ scalar constraints per encoding,
so the space $\mathcal T_k$ of compatible responses has dimension at least
$m^2-3k(m-3)$. Generic families attain this bound with $3m$ free parameters at $k_{\rm tomo}$ and $9$ at $k_{\rm tomo}+1$. At $\kc$ they force
$T=\lambda I_m$ [End Matter and Fig.~\ref{fig:messages}(c)].

To attain the threshold, we partition all but three Pauli axes into Clifford encodings and add three rotated qubits. Analyticity extends this construction to almost every family (End Matter). On an all-to-all register without ancillary qubits, the construction uses $O(\log n)$ layers per encoder and a total of $O(n)$ non-Clifford Pauli rotations with freely chosen angles. At least $2n$ rotations are necessary. With fewer, dephasing about a Pauli that commutes with every rotation preserves all the encoded algebras.

\emph{Two-qubit example.}---Here $\kc=7$. Figure~\ref{fig:messages}(d) compares a $Z\otimes I$ phase flip of probability $1/4$ with depolarization at the same $\lambda=11/15$. Under the phase flip, a bit encoded in $(I_4\pm Q)/4$ is read out with success probability $1$ for $Q=Z\otimes I$ and $3/4$ for $Q=X\otimes I$. Under depolarization, it is $13/15$ in both bases. We construct 7 encoded qubits. The first 6 span all $15$ Pauli directions, so they suffice for state tomography, yet both channels pass all 6 autonomy tests. The seventh qubit is probed with two inputs that have the same logical and spectator marginals but different correlations. Only the phase flip yields a nonzero spectator contrast, $|c_7|\simeq 0.1364$~\cite{SupplementalMaterial}.

\emph{Pauli noise.}---
For a Pauli channel, each physical Pauli $Q_a$ decays at its own rate, $\Phi^\dagger(Q_a)=t_aQ_a$, so $T=\operatorname{diag}(t_a)$. Such channels arise after Pauli twirling, as in noise tailoring~\cite{WallmanEmerson2016}, and their fault probabilities can be learned experimentally~\cite{FlammiaWallman2020,Harper2020,Chen2023}. The full diamond norm of the difference between two Pauli channels equals the $\ell^1$ distance between their fault probabilities~\cite{Sacchi2005}. Among depolarizing channels the unique closest one is $\mathcal D_\lambda$ at the mean rate. For a given family of encodings, let $C$ be the smallest constant satisfying $\|\Phi-\mathcal D_\lambda\|_\diamond\le C\,e(T)$ for all Pauli channels.

\emph{Theorem 2.}---For Haar-almost every encoder, a single encoded qubit is autonomous under a Pauli channel only if the channel is depolarizing. A Clifford-encoded qubit is autonomous under every Pauli channel. Every family of encodings satisfies $C>D/\sqrt{3}$, and a single encoder can achieve the optimal scaling $C=\Theta(D)$.

Let $P=P_1$ be the projector of the encoded qubit. Since $T$ is diagonal, Eq.~\eqref{eq:autonomy} is equivalent to $TP=PT$. In components,
\begin{equation}
 P_{ab}(t_a-t_b)=0 .
 \label{eq:linked_rates}
\end{equation}
Join axes $a,b$ whenever $P_{ab}\ne0$. Equation~\eqref{eq:linked_rates} makes the rates equal along each edge, and hence throughout each connected component. Connectivity is therefore the criterion for forcing depolarization~\cite{SupplementalMaterial}. A Clifford encoder selects three Pauli strings. Its graph has no edges, so autonomy places no constraint on the rates [Fig.~\ref{fig:pauli}(a)]. One rotation $e^{-i\pi XX/8}$ already links two pairs and detects a $Z\otimes I$ phase flip [Fig.~\ref{fig:pauli}(b)]. A generic encoder links all anticommuting pairs, which connects the graph [Fig.~\ref{fig:pauli}(c)]. In this circuit model, a single encoder needs at least $2n$ non-Clifford Pauli rotations to detect all Pauli anisotropies.

The graph also controls approximate autonomy. For one encoding, the squared leakage is
$e^2=\sum_{a<b}|P_{ab}|^2(t_a-t_b)^2=t^\top L t$,
where $L$ is the graph Laplacian with edge weights $|P_{ab}|^2$.
For diagonal responses, the gap of Eq.~\eqref{eq:leak}, denoted $\gamma$, is the square root of the second-smallest eigenvalue of $L$. For a connected graph, $\gamma>0$, and the rate bound $R=\|t-\lambda\mathbf{1}\|_2\le e/\gamma$ is sharp. The Pauli fault probabilities and the Laplacian trace give the respective bounds
\begin{equation}
 \|\Phi-\mathcal D_\lambda\|_\diamond\le e/\gamma,
 \qquad \gamma<\sqrt3/D ,
 \label{eq:pauli_upper}
\end{equation}
so $C\le1/\gamma$ for one encoded qubit. The lower bound of Theorem~2 comes from a channel that transmits a single Pauli axis of least average weighted degree (End Matter). The optimal worst-case ratio of diamond distance to leakage is therefore $\Theta(D)$, a scaling attainable with a single encoding.

The least Pauli weight $w_{\min}=\min_aP_{aa}$ limits the gap through $\gamma^2\le\gamma_w^2=m\,w_{\min}(1-w_{\min})/(m-1)$. In Fig.~\ref{fig:pauli}(d), the median of
$\gamma/\gamma_w$ exceeds $0.98$ at each dimension, while the median of $D\gamma$ falls from $0.41$ at $D=4$ to $0.13$ at $D=32$. Balanced encoders reach $84\%$ and $78\%$ of the finite-$D$ trace bound at $D=4$ and $8$, respectively \cite{SupplementalMaterial}.

\emph{Precision and redundancy.}---
For general channels, the redundancy $\nu=3k-m$ limits the attainable precision. The gap of any family with $k\ge\kc$ satisfies
\begin{equation}
 g^2\le \frac{\nu}{k(m-3)} .
 \label{eq:relation_bound}
\end{equation}
Let $g_*$ be the supremum of $g$ over all families of $k$ distinct encoded qubits.

\emph{Theorem 3.}---Let $D=2^n\ge4$. At the threshold ($\nu=6$), and for $\nu\ge C_0\log D$ with a sufficiently large absolute constant $C_0$, the optimal gap satisfies
\begin{equation}
 (Dg_*)^2 \asymp \frac{\nu}{3k},
 \label{eq:precision}
\end{equation}
with absolute implicit constants~\cite{SupplementalMaterial}.

Thus $Dg_*$ scales as the square root of the fraction of redundant logical axes. The optimal amplification $1/g_*$ therefore scales as $D^2$ at threshold, where only six relations constrain the $m\times m$ response, and as $D$ when a fixed positive fraction of the logical axes is redundant. An explicit family of $k\le4m-2$ Clifford+$T$ encoders, each using at most one $T$ gate, satisfies $kg^2\ge1/6$ and hence $1/g<5D$~\cite{SupplementalMaterial}. At $D=16$, the optimal scaled gap satisfies $0.115<Dg_*<0.265$ at threshold ($k=87$) and $Dg_*>0.998$ at $k=174$ (Fig.~\ref{fig:precision}).

\emph{Worst-case distinguishability.}---
Let $r$ be the ratio of output to input trace distance for any two distinct states. For $g>0$,
\begin{equation}
 \bigl|r-\lvert\lambda\rvert\bigr|
 \le \sqrt{\frac{D}{2}}\,R\le\sqrt{\frac{D}{2}}\,\frac{e(T)}{g}.
 \label{eq:retention}
\end{equation}
The factor $\sqrt{D/2}$ is sharp. For any family of encodings, there is a unital channel with $\lambda=0$ that transmits a bit perfectly in one basis, erases it in another, and satisfies
\begin{equation}
 e(T)\le\sqrt{6/(Dm)} .
 \label{eq:bit-cost}
\end{equation}
Let $Q$ be a Pauli axis of minimum mean projection weight, and let $|u\rangle,|v\rangle$ be orthogonal states in its $+1$ eigenspace. The channel
maps $|u\rangle$ and $|v\rangle$ to $(I_D+Q)/D$
and $(I_D-Q)/D$, respectively, and both superpositions
$(|u\rangle\pm|v\rangle)/\sqrt{2}$ to $I_D/D$ (End Matter).

For every encoding family, bounding the retention error by $e(T)$ thus requires a factor of at least $\sqrt{Dm/6}$. The best families attain this $D^{3/2}$ scaling by Theorem~3 and Eq.~\eqref{eq:retention}. Mixing with the identity extends the same scaling to arbitrarily weak noise. For the full diamond distance to the nearest depolarizing channel, the optimal factor is likewise $\Theta(D^{3/2})$ for unital channels~\cite{SupplementalMaterial} and $\Theta(D)$ for Pauli channels (Theorem~2).

\emph{How many channel uses?}---
Each trial uses the same channel once on a fresh register without
an external quantum reference and records one binary outcome.
Calibrated control settings may depend on past outcomes.
The test must have error probability at most $1/3$ under either hypothesis.

\emph{Theorem 4.}---Consider testing $R=0$ against $R\ge\epsilon$ for $0<\epsilon\le1/4$. The minimax cost is $\widetilde{\Theta}(D^3/\epsilon^2)$ calls for general channels when the binary readout may change between trials, and for Pauli channels with the best fixed encoding. Letting the encoding change between trials lowers the Pauli cost to $\widetilde{\Theta}(D^2/\epsilon^2)$. Both Pauli bounds also hold for alternatives at least $\epsilon$ from every depolarizing channel in full diamond norm. In all three settings, the scaling persists when both hypotheses share the known mean rate $\lambda=1-h$, with $0<h\le1/2$
and $\epsilon\le h/4$~\cite{SupplementalMaterial}.

The tilde denotes logarithmic factors in $D$ and $1/\epsilon$. Autonomy tests attain all three minimax rates for binary readouts. They measure one logical Pauli per trial and compare outcomes for inputs with the same logical marginal.
A fixed encoding probes at least one Pauli axis weakly, since
$\sum_{a=1}^{m}P_{aa}=3$ implies $\min_a P_{aa}\le 3/m$. By switching encoders, one can compare the rates of two anticommuting Paulis directly on each trial (End Matter).

Hidden Pauli rates also give lower bounds for learning Pauli channels without an ancilla~\cite{ChenZhouSeifJiang2022}, and Chen and Gong use them to show how a small quantum memory changes this cost~\cite{PRMinChenGong}. Fawzi \emph{et al.} certify a specified completely depolarizing channel in diamond norm~\cite{PRMinFawzi}. Here the null hypothesis contains every isotropic response, and each trial returns one bit. Within this model the same Pauli test costs order $D^3/\epsilon^2$ calls with a fixed encoding and $D^2/\epsilon^2$ with switching, so the choice of which encoded qubit to read saves one power of $D$.

\emph{Discussion.}---
Forcing isotropy exactly is cheap: two encoded qubits beyond the tomographic count suffice. Forcing it robustly requires redundancy: keeping the leakage amplification of order $D$ needs $\Omega(m)$ independent relations. 

The threshold concerns a family chosen to certify every channel. If instead a particular channel with $R>0$ is fixed first, almost every Haar-random encoding exhibits leakage~\cite{SupplementalMaterial}. For a forcing family, the leakage data determine the anisotropic part $T-\lambda I_m$ itself, with the same amplification $1/g$ as in Eq.~\eqref{eq:master}.

At the same fault probability $0<p<1/2$, the $ZI$ and $XI$ channels, which apply $Z\otimes I$ or $X\otimes I$ with probability $p$, have identical average gate fidelity and unitarity. For anticommuting Paulis $A,B$ with decay rates $t_A,t_B$, the signed readout contrast is $(t_A-t_B)/2$ (End Matter). The pairs $(YI,ZX)$ and $(YI,XX)$ give contrasts $(-p,0)$ for the $ZI$ channel and $(0,-p)$ for the $XI$ channel. Under $ZI$ noise, optimal decoding of an unbiased bit on the first qubit succeeds with probability $1-p$ in the $X$ basis
and with certainty in the $Z$ basis. These probabilities are reversed under $XI$ noise.

Testing the autonomy of enough generic encoded qubits, one at a time, certifies
that every state difference scales by the mean rate $\lambda$, as in a depolarizing
channel. When the noise has a preferred basis, the same tests reveal which messages it protects and which it damages.\begin{acknowledgments}
This work was supported by Hebei University of Science and Technology
under Starting Grant No.~81/1181298.

Claude Opus 5.5 assisted with the plotting code for
Figs.~\ref{fig:messages} and~\ref{fig:pauli}.
\end{acknowledgments}
\newcommand{\SupplementOnlyRefs}{\cite{BarencoGates,BitWatrous,Dankert2009,HasseBound,KeppelerSjodahl2012,Magesan2012,Meckes2019,ReadoutYu,Tropp2012,VanHandel2016}}
\nocite{BarencoGates,BitWatrous,Dankert2009,HasseBound,KeppelerSjodahl2012,Magesan2012,Meckes2019,ReadoutYu,Tropp2012,VanHandel2016}
\bibliographystyle{apsrev4-2}
\makeatletter
\immediate\write\@auxout{\string\citation{apsrev42Control}}
\makeatother
\bibliography{bibliography-control,references}
\raggedbottom
\clearpage
\appendix
\section{End Matter}
\label{sec:endmatter}

\emph{Why two encodings beyond tomography suffice.}
Below $\kc$ the constraint count leaves at least 9 dimensions. Imposing $Tv=0$ for a nonzero $v$ in a tested plane adds at most three constraints, leaving a nonzero singular $T$. The positive definite Choi matrix of complete depolarization permits a small physical perturbation with response proportional to $T$. Perturbing $I_D/D$ along directions inside and outside $\ker T^\top$ gives one pair with identical outputs and another with distinct outputs.

At $k_{\rm tomo}=m/3$ the generic planes form a direct sum, giving $3m$ parameters for their independent $3\times3$ actions. With one more generic plane, removing the three axes of any one encoder leaves a basis. The coordinate projection from the added plane onto each summand is therefore invertible. Conjugating an arbitrary $3\times3$ action on the added plane by these projections determines the other actions, leaving exactly 9 parameters.

To reach $\kc$, take $(m-3)/3$ disjoint Clifford triples, leaving three commuting axes, and add three rotated qubits. For suitable rotations, the first two added planes determine every response block from the block on the uncovered axes. The third restricts this block to the common commutant of two matrices, which consists only of scalars. Clifford equivalence~\cite{DehaeneDeMoor2003} and explicit rotations give a witness for each required minor (Supplemental Material~\cite{SupplementalMaterial}, Sec.~III). After clearing denominators, the minors are analytic on the same connected space of rotations, so they are simultaneously nonzero almost everywhere. The resulting nonzero constraint minor of order $m^2-1$ is analytic in all $\kc$ encoders, proving genericity for the full family.

If deleting a plane leaves a proper span $W$, take $u\ne0$ in the deleted plane and a nonzero functional $\ell$ vanishing on $W$. Then $v\mapsto\ell(v)u$ preserves every plane and is nonscalar, contradicting forcing.

\emph{Pauli connectivity and diamond distance.}
For anticommuting $Q_a,Q_b$, rotate a Pauli plane containing $Q_a$ and orthogonal to $Q_b$ about $iQ_bQ_a$. Then $P_{ba}$ has nonzero derivative at zero angle. By analyticity almost every encoder links all such pairs. Distinct commuting axes share an anticommuting neighbor, so Eq.~\eqref{eq:linked_rates} ties all rates together.

At rate $\lambda$ the identity fault probabilities match and the remaining residuals sum to zero. Shifting the candidate identity probability by $\delta\ne0$ then increases the $\ell^1$ distance, and hence the full diamond distance~\cite{Sacchi2005}, by at least $2|\delta|/m$, so $\mathcal D_\lambda$ is the unique minimizer.

For each encoder, $P^2=P$ gives the degree $d_a=P_{aa}(1-P_{aa})$. Since $\tr P=3$, some axis has family-averaged degree $\bar d_a\le3(m-3)/m^2$. Averaging conjugations by the Paulis commuting with $Q_a$ preserves only that traceless axis. This channel has $e(T)^2=\bar d_a$ and diamond distance $(m-1)/m$ from $\mathcal D_{1/m}$ (Supplemental Material, Sec.~VI). Hence $C\ge(m-1)/\sqrt{3(m-3)}>D/\sqrt3$. If $\bar d_a=0$, no finite $C$ suffices.

\emph{How redundant axes improve precision.}
At threshold, deforming a two-qubit core spreads its six relations across the frame, giving $g\gtrsim1/m$. For $C_0\log D\le\nu\le m$ as in Theorem~3, adding independent Haar encoders to the coordinate triples matches Eq.~\eqref{eq:relation_bound} up to constants with high probability as $D$ grows, and rotated copies extend this scaling to $\nu\ge m$ (Supplemental Material, Secs.~X--XII).

\emph{One message survives and another is erased.}
Choose an axis $Q_a$ with mean projection weight at most $3/m$. For two orthogonal levels $|0\rangle,|1\rangle$ in its positive eigenspace, set $S=|0\rangle\langle0|-|1\rangle\langle1|$. Measure $(I\pm S)/2$ and prepare $(I\pm Q_a)/D$ with the observed sign. The levels give orthogonal outputs, whereas $(|0\rangle\pm|1\rangle)/\sqrt2$ both give $I_D/D$. The adjoint sends $Q_a$ to $S$ and every other traceless Pauli axis to zero. Since $\tr(SQ_a)=0$ and $\tr(S^2)/D=2/D$, the mean rate vanishes and $e(T)^2\le(2/D)(3/m)$, as in Eq.~\eqref{eq:bit-cost}.

\emph{Fixed and switchable binary readouts.}
The alternative channel measures $Q_b$ and prepares $(I\pm\eta Q_a)/D$ with the measured sign, where $0<\eta\le1/2$. The null is complete depolarization for $a\ne b$ and $\mathcal D_{\eta/m}$ for $a=b$. Since $\sum_a P_{aa}=3$, a uniformly hidden Pauli axis has mean projection weight $3/m$ in any fixed readout plane (Supplemental Material, Sec.~XVI).

For Pauli noise, a fixed encoder with $\gamma\gtrsim1/D$ gives mean squared contrast $\gtrsim R^2/D^3$ using randomized preparations. A family with $g\gtrsim1/D$ gives the same bound for general channels. For switchable Pauli readout, sample an edge $\{a,b\}$ uniformly from the anticommutation tree and use its Clifford+$T$ encoder. Measure $(Q_a+Q_b)/\sqrt2$ on the outputs of $(I\pm K)/D$, where $K=(Q_a-Q_b)/\sqrt2$ is orthogonal to the chosen logical plane. The inputs thus have the same logical marginal. Half the difference of the mean readouts is $(t_a-t_b)/2$, with mean square at least $R^2/[12(m-1)]$ over edges. Multiscale repetition gives the upper bounds of Theorem~4 up to logarithms (Supplemental Material, Secs.~XV--XVI).\end{document}


\title{Supplemental Material for ``How Many Autonomous Encoded Qubits Force an Isotropic Noise Response?''}
\maketitle
\paragraph{Conventions.}
The register has $n\ge2$ qubits, $D=2^n$ and $m=D^2-1$. The nonidentity Paulis form an orthonormal basis of the real space $V$ of traceless Hermitian observables for $\langle A,B\rangle=\tr(A^\dagger B)/D$. We write $\|\cdot\|_2$ for the corresponding norm on observables, $\|\cdot\|_F$ for the unnormalized Frobenius norm and $\|\cdot\|$ for the operator norm. A family has $k$ encoded qubits and $\nu=3k-m$ excess logical axes. When the axes span $V$, $\nu$ is the dimension of their relation space. We write $k_{\rm tomo}=m/3$, $\kc=k_{\rm tomo}+2$, $\lambda=\tr T/m$, $R=\|T-\lambda I_m\|_F$, and $x=\nu/m$. The optimal gap over families of $k$ distinct encoded qubits is $g_*(D,k)$. Constants implicit in $\asymp$, $\gtrsim$, $O$ and $\Omega$ are independent of $D$ and $k$.

\section{Response, autonomy and leakage}\label{sec:setup}
Let $F_j\in\mathbb R^{m\times3}$ contain the three logical Pauli axes of encoder $U_j$, and let $P_j=F_jF_j^\top$ project onto their span $V_j$. With the identity first in the Pauli basis, the channel and its adjoint have matrices
\begin{equation}
 [\Phi^\dagger]=\begin{pmatrix}1&b^\top\\0&T\end{pmatrix},
 \qquad [\Phi]=\begin{pmatrix}1&0\\b&T^\top\end{pmatrix}.
 \label{eq:sm_channel_blocks}
\end{equation}
The bias $b$ cancels from every state difference.

\emph{Autonomy.}
Two states have the same logical marginal exactly when their difference is orthogonal to the logical algebra $\mathcal A_j$. For each $Y\in V_j^\perp$ the states $(I_D\pm\delta Y)/D$ realize a difference along $Y$ for sufficiently small $\delta>0$. Equal input marginals therefore give equal output marginals exactly when $\Phi^\dagger(\mathcal A_j)\subseteq\mathcal A_j$, or equivalently
\begin{equation}
 (I_m-P_j)TP_j=0 .
 \label{eq:sm_channel_autonomy}
\end{equation}

\emph{Physical responses.}
A real $m\times m$ matrix $X$ defines a Hermiticity-preserving map $\Delta$ through $[\Delta^\dagger]=\operatorname{diag}(0,X)$. In the convention $J(\Phi)=\sum_{ab}|a\rangle\langle b|\otimes\Phi(|a\rangle\langle b|)$, complete depolarization has Choi matrix $I_{D^2}/D$, so
\begin{equation}
 \Phi_\epsilon=\mathcal D_0+\epsilon\Delta,\qquad
 |\epsilon|<1/(D\|J(\Delta)\|),
 \label{eq:sm_choi_interior}
\end{equation}
is a unital channel with response $\epsilon X$. Every sufficiently small response is therefore physical.

\emph{Leakage and gap.}
On $\mathbb R^{m\times m}$, $\mathcal G_P(X)=(I_m-P)XP$ is an orthogonal projector of rank $3(m-3)$ that annihilates $I_m$. For a family of $k$ planes, define
\begin{equation}
 \mathcal G=\frac1k\sum_j\mathcal G_{P_j},\qquad
 e(X)^2=\langle X,\mathcal GX\rangle_F,\qquad
 g^2=\lambda_{\min}(\mathcal G|_{I_m^\perp}).
 \label{eq:sm_leakage_operator}
\end{equation}
Thus $g>0$ exactly when autonomy of the whole family forces a scalar response. For $g>0$, subtracting the mean response leaves leakage unchanged and gives
\begin{equation}
 R=\|T-\lambda I_m\|_F\le e(T)/g .
 \label{eq:sm_channel_stability}
\end{equation}
For $g>0$ the map from centered responses to the $k$ leakage blocks is injective, so the measured blocks determine $T-\lambda I_m$. If the blocks carry errors of root-mean-square size $\delta$ in the normalization of Eq.~\eqref{eq:sm_leakage_operator}, the least-squares reconstruction of $T-\lambda I_m$ is off by at most $\delta/g$ in Frobenius norm.
The unitarity is $\|T\|_F^2/m=\lambda^2+R^2/m$, and the average fidelity relative to the identity is $[1+(D-1)\lambda]/D$~\cite{WallmanUnitarity2015,Magesan2012}. Scaling a centered unit minimizer of $e$ as in Eq.~\eqref{eq:sm_choi_interior} gives
\begin{equation}
 \sup_{\Phi:e(T)>0}\frac{R}{e(T)}=\frac1g .
 \label{eq:sm_channel_sharp}
\end{equation}
When $g=0$, take a centered unit vector in $\ker\mathcal G$ and add a perturbation with positive leakage. As the perturbation tends to zero, $R/e$ diverges. Equation~\eqref{eq:sm_choi_interior} realizes these ratios with physical channels.

\emph{Measuring leakage.}\label{sec:physical_consequences}
Let $A_{ja}$ be a logical axis and let $\{Y_{jb}\}$ be an orthonormal basis of $V_j^\perp$. The states
\begin{equation}
 \rho_{jb}^{\pm}=(I_D\pm\epsilon Y_{jb})/D,\qquad 0<\epsilon\le1/\|Y_{jb}\|,
 \label{eq:sm_equalmarginals}
\end{equation}
have the same maximally mixed logical marginal. Their scaled logical-output difference gives one leakage entry:
\begin{equation}
 (M_j)_{ba}
 =\frac{\tr[A_{ja}(\Phi(\rho_{jb}^+)-\Phi(\rho_{jb}^-))]}{2\epsilon}
 =\frac{\tr[Y_{jb}\Phi^\dagger(A_{ja})]}{D},
 \label{eq:sm_channel_readout}
\end{equation}
and $\|M_j\|_F=\|(I_m-P_j)TP_j\|_F$.

\emph{Scalar response and state discrimination.}
If $T=\lambda I_m$, linearity gives $\Phi(Y)=\lambda Y+\tr(Y)B$, where $B=\Phi(I_D/D)-\lambda I_D/D$. For orthogonal unit vectors $x,y$, positivity gives $\langle y|B|y\rangle=\langle y|\Phi(|x\rangle\langle x|)|y\rangle\ge0$. Trace preservation gives $\tr B=1-\lambda$. At $\lambda=1$ the channel is the identity. Otherwise
\begin{equation}
 \Phi(Y)=\lambda Y+(1-\lambda)\tr(Y)\tau,
 \label{eq:sm_channel_form}
\end{equation}
with fixed state $\tau=B/(1-\lambda)$. For unital noise, $\tau=I_D/D$. Every trace distance is multiplied by $|\lambda|$.

For distinct states set $A=\rho-\sigma$ and $Y=\Phi(A)-\lambda A$. Then $\|Y\|_F\le R\|A\|_F$. Since $A$ is traceless, its positive and negative parts each have trace $\|A\|_1/2$, so $\|A\|_F\le\|A\|_1/\sqrt2$. With $\|Y\|_1\le\sqrt D\|Y\|_F$ and $g>0$, Eq.~\eqref{eq:sm_channel_stability} gives
\begin{equation}
 \left|\frac{\|\Phi(A)\|_1}{\|A\|_1}-|\lambda|\right|
 \le\sqrt{\frac D2}\,\frac{e(T)}{g}.
 \label{eq:sm_retention_conversion}
\end{equation}

\emph{One random encoder.}\label{sec:random_encoder}
Fix a channel with $R>0$. Its squared leakage $e(T)^2=\|(I-P_U)TP_U\|_F^2$ is real analytic on the connected group $U(D)$. If it vanished identically, $T$ would preserve every encoded plane, contrary to Theorem~1. Its zero set has Haar measure zero and its Haar mean is positive. A Haar-random encoded qubit therefore detects any fixed anisotropic channel almost surely.
\section{Counting encoded qubits}\label{sec:relations}
Each plane imposes at most $3(m-3)$ linear constraints on $T$. For $k<\kc$, the compatible response space therefore satisfies
\begin{equation}
 \dim\bigcap_j\ker\mathcal G_{P_j}
 \ge m^2-3k(m-3)\ge9 .
 \label{eq:sm_channel_lowercount}
\end{equation}
For a nonzero $v$ in a tested plane, adding $Xv=0$ removes at most three further dimensions and leaves a nonzero singular response. Equation~\eqref{eq:sm_choi_interior} makes a small multiple physical. Directions inside and outside $\ker X^\top$ then supply state pairs with different trace-distance retention, although every tested qubit is autonomous.

\begin{lemma}[Minimum family for state tomography]\label{lem:tomography_minimum}
The least number of encoded-qubit marginals that determine every state is
\begin{equation}
 k_{\rm tomo}=m/3,
 \label{eq:sm_tomography_minimum}
\end{equation}
and Haar-almost every family of this size attains it.
\end{lemma}
The logical axes must span $V$, since a nonzero traceless $Y$ orthogonal to their span gives distinct states $(I_D\pm\epsilon Y)/D$ with identical marginals for small $\epsilon$. This proves $3k\ge m$. Use the $(m-3)/3$ coordinate qubits of Sec.~\ref{sec:exact} and append the first base encoder of Table~\ref{tab:rotations}, extended by identities on the other qubits. Its projection onto the three missing directions has determinant $113799168/5^{14}$, so the resulting square frame is invertible. Its determinant is analytic in the encoders and hence nonzero for Haar-almost every family of this size.

\emph{Tomography after deleting one qubit.}
Suppose that removing plane $j$ from a forcing family leaves a proper span $W$. Choose $u\ne0$ in plane $j$ and a nonzero functional $\ell$ vanishing on $W$. Then the rank-one response $X(v)=\ell(v)u$ is nonscalar and preserves every plane of the original family, contradicting forcing. Deleting any one member therefore leaves a tomographically complete family. At threshold the remaining $\kc-1$ planes admit nonscalar responses by Eq.~\eqref{eq:sm_channel_lowercount}.

\section{Encoders at the threshold}\label{sec:exact}
The threshold construction uses $p=(m-3)/3$ Clifford encoders and three rotated encoders. The Clifford planes cover all but three commuting Pauli axes. The rotated planes then determine every response block and restrict the remaining block to scalar multiples of the identity.

\emph{Pauli near-partition.}
For $v=(x,z)\in\mathbb F_2^{2n}$, write $Q_v=i^{x\cdot z}X^xZ^z$, taking the exponent over the integers and listing the highest wire first. The tables index $Q_v$ by $\sum_{i=0}^{n-1}2^i(x_i+2^nz_i)$. Two Paulis anticommute exactly when their binary symplectic form $[v,w]=x\cdot z'+z\cdot x'$ equals one. A line is a triple $\{u,v,u+v\}$ of nonzero labels. It spans a Clifford qubit plane when $[u,v]=1$ and a commuting plane when $[u,v]=0$.

Table~\ref{tab:basecoords} gives the two-qubit partition. To extend it from $r$ to $r+1$ qubits, let $G$ be block diagonal with blocks
\[
 G_2=\begin{pmatrix}0&1\\1&1\end{pmatrix},\qquad
 G_3=\begin{pmatrix}0&0&1\\1&0&1\\0&1&0\end{pmatrix},
\]
using only $2\times2$ blocks for even $r$ and one $3\times3$ block for odd $r$. The map $\Gamma(x,z)=(Gx,G^\top z)$ satisfies $[w,\Gamma w]=0$, with $\Gamma$ and $I+\Gamma$ invertible. For a symplectic pair $\alpha,\beta$ on the new wire, the $4^r$ qubit lines
\[
 \{(w,\alpha),(\Gamma w,\beta),(w+\Gamma w,\alpha+\beta)\},\qquad w\in\mathbb F_2^{2r},
\]
cover every label with nonzero last component exactly once. Together with the old partition, this gives $p$ coordinate qubits and one commuting line in every dimension. Denote their spans by $V_1,\ldots,V_p$ and $V_0$.

\begin{table}[ht]
\caption{The two-qubit Pauli partition. Strings list the high-order tensor factor first.}\label{tab:basecoords}
\begin{ruledtabular}
\begin{tabular}{lll}
Space & Labels & Pauli strings\\ \hline
$V_1$ & $(1,4,5)$ & $IX,IZ,IY$\\
$V_2$ & $(2,9,11)$ & $XI,ZX,YX$\\
$V_3$ & $(6,8,14)$ & $XZ,ZI,YZ$\\
$V_4$ & $(7,10,13)$ & $XY,YI,ZY$\\
$V_0$ & $(3,12,15)$ & $XX,ZZ,YY$
\end{tabular}
\end{ruledtabular}
\end{table}

\emph{Three additional planes.}
A response preserving every coordinate plane acts on $V_i$ through a $3\times3$ matrix $A_i$. Its remaining blocks are $B_i:V_0\to V_i$ and $A_0:V_0\to V_0$. Write an additional plane that projects invertibly onto $V_0$ as the graph
\[
 \{(Y_{1j}y,\ldots,Y_{pj}y,y):y\in\mathbb R^3\}.
\]
Its invariance is equivalent to
\begin{equation}
 A_iY_{ij}+B_i=Y_{ij}A_0\qquad(1\le i\le p).
 \label{eq:sm_graph_invariance}
\end{equation}
For three such planes, define $\Delta_i=Y_{i2}-Y_{i1}$ and $\Delta_i'=Y_{i3}-Y_{i1}$. Subtracting the equation for the first plane gives
\[
 A_i\Delta_i=\Delta_iA_0,\qquad A_i\Delta_i'=\Delta_i'A_0.
\]
If every $\Delta_i$ is invertible, then $A_i=\Delta_iA_0\Delta_i^{-1}$ and $A_0$ commutes with $M_i=\Delta_i^{-1}\Delta_i'$. When the common commutant of $M_1$ and $M_2$ consists only of scalars, all $A_i$ equal $\lambda I_3$ and all $B_i$ vanish. Hence the full response is $\lambda I_m$.

\emph{Physical witnesses and genericity.}
The frame of encoder $U_j$ has columns $U_jQ_aU_j^\dagger$ for $a=1,2^n,1+2^n$, the $X,Z,Y$ axes on the least significant wire. A wire permutation identifies this qubit with the reference qubit of the Letter. If $Z_j$ and $X_{ij}$ are its row blocks in $V_0$ and $V_i$, then $Y_{ij}=X_{ij}Z_j^{-1}$. We need $\det Z_j\ne0$, $\det\Delta_i\ne0$, and a scalar common commutant for $M_1,M_2$.

Each $\det\Delta_i$ depends only on the coordinate qubit line and the commuting line. Their binary planes intersect at zero. Basis changes within the two planes reduce their $2\times2$ symplectic cross-pairing to rank zero, one or two. By Witt's theorem an isometry between two pairs of the same rank extends to a symplectic map of the full space, implemented by a Clifford up to Pauli signs~\cite{DehaeneDeMoor2003}. Signed row permutations preserve nonvanishing. With three qubits, use the commuting line $(8,16,24)$ and the qubit lines $(4,32,36)$, $(1,40,41)$ and $(1,10,11)$ for the three ranks in order. Tensoring these witnesses with identities covers $n\ge3$. The base in Table~\ref{tab:basecoords} covers $n=2$ and gives the tests on $Z_j$ and the common commutant in every dimension.

In Table~\ref{tab:rotations}, $(a_1,\ldots,a_l)$ denotes $U(a_l)\cdots U(a_1)$ with $U(a)=(3I+4iQ_a)/5$. Conjugation fixes commuting Paulis and sends an anticommuting $Q_v$ to $(-7Q_v+24iQ_aQ_v)/25$. Table~\ref{tab:witness} gives the determinant residues. For the base, represent $A\mapsto(AM_1-M_1A,AM_2-M_2A)$ by listing each matrix in row order, with indices starting at zero. Its minor on rows $(0,1,2,3,4,5,9,10)$ and columns $(0,\ldots,7)$ is $170$ modulo $1009$. The map has rank at least eight. Since the domain has dimension nine and the kernel contains $I_3$, the common commutant consists of scalars.

\begin{table}[ht]
\caption{Rotation sequences of the witnesses.}\label{tab:rotations}
\begin{ruledtabular}
\begin{tabular}{lcl}
Witness & Encoder & Pauli labels $(a_1,\ldots,a_l)$\\ \hline
Base & $U_1$ & $(14,15,9,8)$\\
 & $U_2$ & $(12,2,1,15,3,10)$\\
 & $U_3$ & $(3,14,1,7,2)$\\
Rank 0 & $U_1$ & $(12,17,45,6,58)$\\
 & $U_2$ & $(21,39,62,11,15,29)$\\
Rank 1 & $U_1$ & $(24,39,22,23,49,17)$\\
 & $U_2$ & $(41,17,46,22)$\\
Rank 2 & $U_1$ & $(49,62,59,48,37,1,39)$\\
 & $U_2$ & $(62,43,50,39)$
\end{tabular}
\end{ruledtabular}
\end{table}

\begin{table}[ht]
\caption{Determinants modulo $1009$, ordered by encoder and line.}\label{tab:witness}
\begin{ruledtabular}
\begin{tabular}{cll}
Witness & $\det Z_j$ & $\det\Delta_i$\\ \hline
Base &$(834,757,757)$&$(29,793,791,804)$\\
Rank 0 &$(672,607)$&$993$\\
Rank 1 &$(378,67)$&$159$\\
Rank 2 &$(487,67)$&$83$
\end{tabular}
\end{ruledtabular}
\end{table}

After clearing denominators, each test is a real analytic function on the same connected manifold $U(D)^3$. Each has a nonzero witness, so the union of their zero sets has Haar measure zero. Together with the coordinate planes, such a triple gives constraint rank $m^2-1$. A minor of this order is thus nonzero somewhere on $U(D)^{\kc}$ and hence nonzero almost everywhere. For $k>\kc$ the first $\kc$ encoders already force a scalar response.

\section{The two-qubit example of Fig.~1}
\label{sec:phase_flip_example}
The phase flip $\Phi_{\rm f}(\rho)=\tfrac34\rho+\tfrac14Z_1\rho Z_1$, with $Z_1=Z\otimes I_2$, has rate $1$ on the seven Pauli axes commuting with $Z_1$ and rate $1/2$ on the eight anticommuting axes. Its mean rate is $\lambda=11/15$. Two equally likely messages $(I_4\pm Q)/4$ have optimal output discrimination probability $(1+|t_Q|)/2$. The phase flip gives $1$ for $Q=Z_1$ and $3/4$ for $Q=X\otimes I_2$, whereas $\mathcal D_{11/15}$ gives $13/15$ in both bases.

Start from the triples $(IX,IZ,IY)$ for qubits 1 and 7, $(XI,ZI,YI)$ for qubit 2, and $(XI,ZX,YX)$ for qubits 3 to 6. Conjugate triple $j$ by $U_j=(5I_4-K_j)^{-1}(5I_4+K_j)$. The seven matrices $K_j$ are given in \path{data/phase_flip/encoders.json}. The first six unitaries commute with $Z_1$, so their logical axes remain within its rate eigenspaces. All six qubits are autonomous under both channels, and their axes span all 15 Pauli directions.

For qubit 7, let $H=U_7(YX)U_7^\dagger$, $A=U_7(IZ)U_7^\dagger$, and $\rho^\pm=(I_4\pm H)/4$. Both states have maximally mixed logical and spectator marginals; only their correlations differ. Define
\[
 c_7(\Phi)=\tfrac12\tr\!\left[A\bigl(\Phi(\rho^+)-\Phi(\rho^-)\bigr)\right].
\]
Then
\[
 c_7(\Phi_{\rm f})=-\frac{77410013810}{567521020443},
 \qquad c_7(\mathcal D_{11/15})=0 .
\]
The magnitude $|c_7|\simeq0.1364$ is the largest contrast over the three logical readouts and twelve complementary Pauli probes.

Exact elimination modulo $1009$ and $1013$ shows that every six of the seven frames have rank $15$, while all seven autonomy constraints have rank $224=m^2-1$. Thus every six-qubit subfamily is tomographically complete, and the seven together force a scalar response. The scaled gap is $4g\simeq0.0141$.
\section{Circuit resources}\label{sec:advance_magic}
Encoders are ancilla-free circuits on an all-to-all register. A layer consists of disjoint one- and two-qubit gates. We count the largest depth of an encoder and the total number of non-Clifford Pauli rotations in the family, with arbitrary rotation angles.

\emph{Depth $O(\log n)$ and $O(n)$ rotations suffice.}
Any $s\le4$ independent Pauli labels on $q$ wires can be compressed by a Clifford onto at most $\min(4,q)$ wires. If their symplectic form has rank $2h$, they span $h$ symplectic pairs and $s-2h$ further commuting labels, which a symplectic basis places on $s-h$ wires. Arrange the wires as the leaves of a balanced binary tree and reserve the first $\min(4,q)$ wires of each subtree. To compress four given labels, compress their restrictions to the two children and then apply a Clifford to the at most eight active wires. The merges on one level are disjoint, so the compression $C$ has $O(\log n)$ layers of disjoint merges. Each merge acts on at most eight wires and has bounded depth in one- and two-qubit gates. Order the leaves so that the reference qubit and the two-qubit core of Sec.~\ref{sec:exact} are among the active wires of the root.

Each coordinate qubit of Sec.~\ref{sec:exact} has a Clifford encoder. After computing the compression $C$ of its generating pair, map the reference pair to the compressed pair at the root and apply $C^{-1}$. For each of the three additional encoders take a variable unitary at the root followed by variable gates on the wire sets of the inverse merges. Each gate acts on at most eight wires and decomposes into a bounded number of CNOTs and one-qubit rotations~\cite{BarencoGates}. The family therefore has depth $O(\log n)$ and $O(n)$ non-Clifford rotations.

The parameter space of this layout is a connected product of unitary groups, and each condition of Sec.~\ref{sec:exact} is realized on it. For the test on $\Delta_i$, compress the two generators of line $i$ and the two generators of the commuting line. For $n\ge3$ the root holds three wires for the witness of the appropriate rank. Follow the witness by a Clifford that carries its two lines to the compressed pair, and undo $C$. The determinant is transferred to the required rows up to signed permutations. The conditions on $Z_j$ and on the common commutant use the padded two-qubit base with identity merges, which also covers $n=2$. Every condition is thus nonzero somewhere on the same manifold, and almost every choice of the three variable circuits satisfies all of them.

\emph{A forcing family requires at least $2n$ non-Clifford rotations.}
Move the Clifford gates of each encoder to its input. Every remaining rotation is then about a Pauli. If the family contains fewer than $2n$ rotations, their labels span a proper subspace of $\mathbb F_2^{2n}$ and some $v\ne0$ is symplectically orthogonal to all of them. Conjugation by $Q_v$ commutes with every rotation and acts by signs on each initial Clifford plane, so it preserves every encoded algebra. Under the dephasing channel $(\mathrm{Id}+\Ad_{Q_v})/2$ all encoded qubits are autonomous, while the Paulis that anticommute with $Q_v$ are erased and the others are preserved.
\section{Pauli noise and one encoded qubit}
\label{sec:pauli_orbit}
Label the Paulis by $a\in\mathbb F_2^{2n}$, with $Q_0=I_D$ and $t_0=1$. Write $t=(t_a)_{a\ne0}$ for the nonidentity rates. Conjugation by $Q_v$ multiplies $Q_a$ by $\chi_a(v)=(-1)^{[a,v]}$. A Pauli channel and its adjoint are diagonal in this basis:
\begin{equation}
 \Phi(X)=\sum_v p_vQ_vXQ_v,\qquad
 p_v=D^{-2}\sum_a\chi_a(v)t_a .
 \label{eq:sm_cov_pauli_weights}
\end{equation}
Write the rows of the encoder frame $F$ as $f_a^\top$. The plane projector $P=FF^\top$ has entries $P_{ab}=f_a^\top f_b$, and the Pauli weights $w_a=P_{aa}$ sum to three.

\emph{Connectivity criterion.}
Since $T=\operatorname{diag}(t)$, autonomy is equivalent to $[T,P]=0$, or $P_{ab}(t_a-t_b)=0$. Join $a,b$ when $P_{ab}\ne0$ and assign edge weight $P_{ab}^2$. The weighted graph Laplacian is $L=\operatorname{diag}(w)-P\circ P$, and
\begin{equation}
 e(T)^2=\tfrac12\|[T,P]\|_F^2
 =\sum_{a<b}P_{ab}^2(t_a-t_b)^2
 =t^\top Lt
 \ge\gamma^2\|t-\lambda\mathbf1\|_2^2 ,
 \label{eq:sm_cov_gap}
\end{equation}
where $\gamma^2$ is the second-smallest eigenvalue of $L$. Zero leakage makes the rates constant on each connected component. A connected graph therefore forces
\begin{equation}
 \mathcal D_\lambda(X)=\lambda X+(1-\lambda)\tr(X)I_D/D,
 \qquad -1/m\le\lambda\le1 .
 \label{eq:sm_cov_depolarizing}
\end{equation}
At complete depolarization every fault has probability $D^{-2}$, so Eq.~\eqref{eq:sm_cov_pauli_weights} permits all sufficiently small rate perturbations. If the graph is disconnected, such rates can differ between components at zero leakage. If it is connected, taking the perturbation along a $\gamma^2$ eigenvector attains equality in Eq.~\eqref{eq:sm_cov_gap}.

For anticommuting $A,B$, choose a Pauli that anticommutes with $A$ outside the binary span of $A,B$. It generates with $A$ a plane orthogonal to $B$. Rotating this plane about $iBA$ gives a nonzero derivative of $P_{ba}$ at zero angle. Thus almost every encoder links all anticommuting pairs by analyticity. Distinct commuting Paulis share an anticommuting neighbor, so the graph is connected. A Clifford plane has no off-diagonal overlaps and imposes no constraint on Pauli rates.

\begin{lemma}[Closest depolarizing channel]\label{lem:nearest_pauli}
For a Pauli channel with mean rate $\lambda$, $\mathcal D_\lambda$ is the unique closest depolarizing channel in diamond norm. For a connected encoder,
\begin{equation}
 \|\Phi-\mathcal D_\lambda\|_\diamond
 \le\frac{\sqrt m}{D}\frac{e(T)}{\gamma}.
 \label{eq:sm_cov_diamond}
\end{equation}
\end{lemma}
The diamond distance is the $\ell^1$ distance between Pauli fault probabilities \cite{Sacchi2005}. Choosing $z=\lambda$ matches the identity probability and leaves nonidentity residuals $r_a$ of sum zero. A shift $\delta$ in the identity probability adds $|\delta|$ to the distance and changes each other residual by $\delta/m$. At least one such residual has the same weak sign as $\delta$, so the distance increases by at least $2|\delta|/m$. Parseval gives $\|r\|_2=\|t-\lambda\mathbf1\|_2/D$, and Cauchy--Schwarz yields Eq.~\eqref{eq:sm_cov_diamond}.

\emph{Bounds on the graph gap.}
Because $P^2=P$, vertex degree is $d_a=w_a(1-w_a)$. Hence
\begin{equation}
 \gamma^2\le\frac{\tr L}{m-1}
 \le\frac{3(m-3)}{m(m-1)}<\frac3{D^2}.
 \label{eq:sm_cov_ceiling}
\end{equation}
Testing the centered indicator of one vertex gives the sharper weight bound
\begin{equation}
 \gamma^2\le\gamma_w^2
 =\frac{m}{m-1}w_{\min}(1-w_{\min}),
 \qquad w_{\min}=\min_aw_a .
 \label{eq:sm_min_degree}
\end{equation}
Let $Q$ have weight $w_{\max}$ and complete its normalized logical projection $A$ to a logical triple $A,H_1,H_2$. Then $E=Q-\sqrt{w_{\max}}A$ has $\|E\|_2^2=1-w_{\max}$. The squared norm of $A$ outside $Q$ is $1-w_{\max}$. For $H=H_1,H_2$, its component commuting with $Q$ has norm $\|\{Q,H\}\|_2/2=\|\{E,H\}\|_2/2\le\|E\|_2$. The $D^2/2-2\ge6$ nonidentity Paulis other than $Q$ commuting with $Q$ thus have total weight at most $3(1-w_{\max})$. Hence $w_{\min}\le(1-w_{\max})/2$. For any weight $w$, the degree difference $(w-w_{\min})(1-w-w_{\min})$ is nonnegative.

\emph{Diamond-distance amplification.}
Choose a vertex $a$ of least degree. With probability $p\in(0,1]$ apply a uniformly chosen Pauli commuting with $Q_a$, and with probability $1-p$ apply $\mathcal D_0$. Its rates are $t_b=p\delta_{ab}$, so $\lambda=p/m$ and $e=p\sqrt{d_a}$. Its fault probabilities give diamond distance $p(m-1)/m$ from $\mathcal D_\lambda$. If $d_a=0$, this positive distance at zero leakage rules out any finite constant $C$. For $d_a>0$, the bound $d_a\le3(m-3)/m^2$ gives
\begin{equation}
 \frac{\|\Phi-\mathcal D_\lambda\|_\diamond}{e}
 =\frac{m-1}{m\sqrt{d_a}}
 \ge\frac{m-1}{\sqrt{3(m-3)}}>\frac{D}{\sqrt3}.
 \label{eq:sm_cov_cost}
\end{equation}
Together with Lemma~\ref{lem:nearest_pauli}, this proves $C>D/\sqrt3$ for every encoder. The construction in Sec.~\ref{sec:flat_all} gives $\gamma\gtrsim1/D$ and hence the optimal order $C=\Theta(D)$.

For a family, replace $L$ by $\bar L=k^{-1}\sum_jL_j$. Its least degree satisfies $\bar d_a\le3(m-3)/m^2$. The same channel gives $C=\infty$ if $\bar d_a=0$ and $C>D/\sqrt3$ otherwise. The averaged Laplacian also obeys Eq.~\eqref{eq:sm_cov_ceiling}.

\emph{Explicit encoders.}
Start from logical axes $XI,YI,ZI$ and apply $e^{-i\pi XX/8}$. The axes become $XI$, $(YI+ZX)/\sqrt2$ and $(ZI-YX)/\sqrt2$, producing links $\{YI,ZX\}$ and $\{ZI,YX\}$, each with $|P_{ab}|=1/2$. A $ZI$ flip of probability $1/4$ gives rates $1/2$ and $1$ across both links and leakage $e=1/(2\sqrt2)$.

For the rational encoder in Fig.~2(c), let $v=(1,2,3,i)^\top$ and $U=I_4+(i-1)vv^\dagger/15$ act on the reference algebra $I_2\otimes M_2$. The figure exchanges the tensor factors. All $105$ pairs of axes are linked, $\min_{a<b}|P_{ab}|=1/1875$, and $900F$ is integral. Exact elimination gives $14$ positive pivots for
\begin{equation}
 \Delta^\top\!\left[25^2\{\operatorname{diag}(K\mathbf1)-K\}
 -6^2\,900^4I_{15}\right]\Delta
 \label{eq:sm_cov_witness}
\end{equation}
with $K=(900^2P)\circ(900^2P)$ and $\Delta=[\delta_1-\delta_{15},\ldots,\delta_{14}-\delta_{15}]$. Thus $\gamma>6/25$, with numerical value $\gamma\simeq0.24408$.

The balanced encoders of Sec.~\ref{sec:finite-pauli-designs} reach $84\%$ and $78\%$ of the universal ceiling at $D=4$ and $8$, respectively.
\section{Balanced encoders on two and three qubits}
\label{sec:finite-pauli-designs}
\emph{Uniform weights.}
If all Pauli weights are equal, $w_a=3/m$, then
\begin{equation}
 \gamma^2\le\frac{9}{5m}.
 \label{eq:sm_uniform_weight}
\end{equation}
Indeed $L=(3/m)I_m-P\circ P$, and $P\circ P$ is the Gram matrix of the symmetric $3\times3$ matrices $f_af_a^\top$, so its rank is at most six. Its trace is $9/m$ and its eigenvalue on $\mathbf1$ is $3/m$. On $\mathbf1^\perp$ there are at most five nonzero eigenvalues and their sum is $6/m$. At least one is therefore at least $6/(5m)$, giving the stated upper bound on the gap of $L$.

\emph{Two qubits: the icosahedral encoder.}
Let $\varphi=(1+\sqrt5)/2$ and let the six columns of
\[
 W=\frac{1}{\sqrt{2(\varphi+2)}}
 \begin{pmatrix}
 0&0&1&-1&\varphi&\varphi\\
 1&-1&\varphi&\varphi&0&0\\
 \varphi&\varphi&0&0&1&-1
 \end{pmatrix},\qquad WW^\top=I_3 ,
\]
point along the six axes through opposite vertices of an icosahedron. The fifteen frame rows are the cross products $f_{ij}=W_{:,i}\times W_{:,j}$, $i<j$. They are realized by Majorana operators $c_{2j-1}=Z_1\cdots Z_{j-1}X_j$ and $c_{2j}=Z_1\cdots Z_{j-1}Y_j$, $j=1,2,3$, on three qubits. On the even-parity sector, identified with two qubits by $|a,b\rangle\mapsto|a,b,a\oplus b\rangle$, the fifteen bilinears $-ic_ic_j$ are the fifteen nonidentity Paulis up to sign and order. Indeed, the six Majorana labels are independent and their product equals parity up to phase. Two distinct bilinears differ by a product of two or four Majoranas, so they remain distinct up to sign on the even-parity sector. The operators $\tilde c_b=\sum_jW_{bj}c_j$, $b=1,2,3$, are anticommuting Hermitian involutions. The triple $-i\tilde c_2\tilde c_3$, $-i\tilde c_3\tilde c_1$, $-i\tilde c_1\tilde c_2$ preserves parity, obeys the Pauli multiplication rules, and has coordinates $f_{ij}$ on $-ic_ic_j$. The first axis exchanges the two eigenspaces of the third. Each eigenspace therefore has dimension two, as required for a logical Pauli triple on this four-dimensional sector.

With $\varphi^2=\varphi+1$ one finds
\[
 \|f_{ij}\|^2=\tfrac15,\qquad
 \sum_{i<j}f_{ij,a}f_{ij,b}f_{ij,c}f_{ij,d}
 =\tfrac1{25}(\delta_{ab}\delta_{cd}+\delta_{ac}\delta_{bd}+
                \delta_{ad}\delta_{bc}).
\]
All fifteen weights equal $1/5$. The second identity says that the frame operator of the matrices $f_{ij}f_{ij}^\top$ sends a symmetric $X$ to $[(\tr X)I_3+2X]/25$. The nonzero eigenvalues of $P\circ P$ are therefore $1/5$ once and $2/25$ five times, and
\[
 \gamma^2=\tfrac15-\tfrac2{25}=\tfrac3{25},
\]
which saturates Eq.~\eqref{eq:sm_uniform_weight} at $m=15$.

\emph{Three qubits.}
The matrix $K$ in \path{data/pauli_designs/D8_encoder.json} is skew-Hermitian with Gaussian integer entries. Its Cayley transform $U=(2^{26}I_8-K)^{-1}(2^{26}I_8+K)$ gives a scaled gap in the interval
\[
 1.341258<8\gamma<1.341261 .
\]
For the lower bound, add $\mathbf1\mathbf1^\top/63$ to $L$, which lifts the zero eigenvalue to one, and subtract $(1.341258/8)^2I_{63}$. After preconditioning with a numerical inverse Cholesky factor the result differs from the identity by a matrix of maximum absolute row sum below one, which proves positivity. For the upper bound a Rayleigh quotient of $L$ on $\mathbf1^\perp$ lies below $(1.341261/8)^2$. Both use $192$-bit outward-rounded arithmetic.
\section{Haar average of one encoded plane}
\label{sec:sm_haar_spectrum}
Sections~\ref{sec:flat_all} and~\ref{sec:random_completion} use the Haar average of the leakage projector of Eq.~\eqref{eq:sm_leakage_operator}. Irreducibility of the adjoint action gives $\mathbb EP_U=(3/m)I_m$, hence
\begin{equation}
 \overline{\mathcal G}(X)=\mathbb E\,(I-P_U)XP_U
 =\frac3mX-\overline{\mathcal P}(X),\qquad
 \overline{\mathcal P}(X)=\mathbb E\,P_UXP_U .
 \label{eq:sm_haar_operator}
\end{equation}
Vectorizing responses identifies their complexification with $V_{\C}\otimes V_{\C}$, $V_{\C}=\mathfrak{sl}_D(\C)$, on which $\overline{\mathcal P}=\mathbb E(P_U\otimes P_U)$ commutes with the conjugation action and with the exchange of the two factors. This representation decomposes into a singlet, a symmetric and an antisymmetric adjoint, and four further irreducible sectors $R_{ab}$, $a,b=\pm$~\cite{KeppelerSjodahl2012}. To describe them, embed $V_{\C}\otimes V_{\C}$ in $(\C^D)^{\otimes2}\otimes((\C^D)^*)^{\otimes2}$, let $S_u$ and $S_\ell$ exchange the two upper and the two lower indices, and put $\Pi_{ab}=(I+aS_u)(I+bS_\ell)/4$. Then $R_{ab}$ is the contraction-free part of $\Pi_{ab}$. The other parts of $\Pi_{aa}$ are a singlet and an adjoint, and of $\Pi_{a,-a}$ an adjoint, so
\[
 \dim R_{aa}=\Bigl[\frac{D(D+a)}2\Bigr]^2-D^2,\qquad
 \dim R_{+-}=\dim R_{-+}=\frac{m(D^2-4)}4 ,
\]
and all dimensions add up to $m^2$. The two adjoints have opposite exchange parity, so Schur's lemma applies to every sector: the eigenvalue of $\overline{\mathcal P}$ is the weight of $P\otimes P$ in the sector divided by its dimension.

Work with the unnormalized Hilbert--Schmidt inner product and the orthonormal axes $A_i=(\sigma_i\otimes I_{D/2})/\sqrt D$, so that $P\otimes P=\sum_{ij}|A_i\otimes A_j\rangle\langle A_i\otimes A_j|$. Since $A_i^2=I/D$, each single swap has expectation $1/D$ in $A_i\otimes A_j$ and the double swap has expectation $\delta_{ij}$. Summing over $i,j$,
\[
 \tr[\Pi_{ab}(P\otimes P)]=\frac{9+9(a+b)/D+3ab}4 .
\]
The contractions are subtracted as follows. The singlet is $(\mathsf{Swap}-I/D)/\sqrt m$ and has weight $3/m$, of which the fraction $(1-a/D)/2$ lies in $\Pi_{aa}$. The symmetric adjoint has weight zero, because the traceless part of every anticommutator $\{A_i,A_j\}$ vanishes. The projection onto the antisymmetric adjoint is $\mathsf{br}^\dagger\mathsf{br}/(2D)$ with $\mathsf{br}(Y\otimes Z)=[Y,Z]$, by the Casimir identity $\sum_j[E_j,[E_j,Y]]=2DY$ for an orthonormal Hermitian basis $E_j$ of $V$. Its weight is $\sum_{ij}\|[A_i,A_j]\|_F^2/(2D)=12/D^2$, shared equally by $\Pi_{+-}$ and $\Pi_{-+}$, which Hermitian conjugation exchanges. The contraction weights are therefore $3(1-a/D)/(2m)$ in $\Pi_{aa}$ and $6/D^2$ in $\Pi_{a,-a}$. Subtracting them and dividing by the dimensions gives the eigenvalues of $\overline{\mathcal P}$,
\begin{align}
 \beta_{aa}&=\frac{6(2D+a)(D+2a)}{D^3m(D+3a)},&
 \beta_{+-}=\beta_{-+}&=\frac6{D^2m},\nonumber\\
 \beta_{\rm A}&=\frac{12}{D^2m},&
 \beta_{\rm S}&=0 ,
 \label{eq:sm_haar_eigenvalues}
\end{align}
on $R_{aa}$, $R_{\pm\mp}$ and the antisymmetric and symmetric adjoints. On the singlet, $\beta=3/m$. Thus $\overline{\mathcal G}$ has eigenvalue zero on the scalar response and eigenvalues $3/m-\beta$ on the centered sectors. Their sum with multiplicities is $3(m-3)$, the rank of a single leakage projector.
\section{An encoder with graph gap of order \texorpdfstring{$1/D$}{1/D} in every dimension}
\label{sec:flat_all}
We construct an encoder with $\gamma\ge c/D$ for an absolute $c>0$. For large $D$ the Paulis that shift the computational basis form a complete graph with edge weights of order $D^{-4}$. Each diagonal Pauli has total edge weight at least a constant times $D^{-2}$ into this graph.

\emph{Core encoders.}
Fix $d=64$, $\hat\sigma_j=\sigma_j\otimes I_{d/2}$ and $v_U(C)=(\tr[C U\hat\sigma_jU^\dagger])_{j=1}^3$. Let $\mathcal C$ consist of four-tuples $\mathbf C=(C_l)$ of zero-diagonal Hermitian Gaussian-integer matrices of size $d$ with $\sum_l C_l\ne0$ and real and imaginary entries bounded in magnitude by $4d$. For $v(\mathbf C)=\sum_l v_{U_l}(C_l)$ we choose four unitaries satisfying the following finite set of conditions:
\begin{enumerate}
\item $v(\mathbf C)^\top v(\mathbf C')\ne0$ for every $\mathbf C,\mathbf C'\in\mathcal C$, including equal tuples;
\item the four $v_{U_l}(\operatorname{diag}s)$ are affinely independent for every nonconstant $s\in\{\pm1\}^d$;
\item deleting the diagonal entries is injective on each logical plane.
\end{enumerate}

For (1), take all four unitaries equal. For any nonzero traceless Hermitian $C,C'$, the pairing $v_U(C)^\top v_U(C')$ is proportional to the inner product of the plane projector $P_U$ with $C\odot C'=(C\otimes C'+C'\otimes C)/2$. The reference projector has a singlet component and no symmetric-adjoint component, as $\sum_j\hat\sigma_j^2=3I$. On the two logical factors $\sum_j\hat\sigma_j\otimes\hat\sigma_j$ acts as $2\,\mathsf{Swap}-I$. Both eigenvalues occur in each full swap-parity sector, by choosing the spectator parity. The $R_{++}$ and $R_{--}$ components are therefore both nonzero. By the decomposition in Sec.~\ref{sec:sm_haar_spectrum}, the orbit spans all sectors except the symmetric adjoint. As a map on traceless observables, $C\odot C'$ has rank at most two. Every nonzero symmetric-adjoint element $X\mapsto HX+XH-(2/d)\tr(HX)I$ has higher rank because at least two pairs $i<j$ of eigenvalues of $H$ have nonzero sum, each furnishing two off-diagonal eigenvectors. Thus some $U$ gives a nonzero pairing. Apply this to the two matrix sums of the tuples.

For (2), the traceless part of $\operatorname{diag}s$ has a nonzero projection onto some logical plane by the same pairing argument with $C=C'$. Logical rotations place this vector at four affinely independent points on its sphere. For (3), an inverse Bell-basis change on the logical qubit and one spectator gives each computational basis state zero expectation of all three axes. Each test is the nonvanishing of a polynomial that is nonzero somewhere on $U(d)^4$. The tests therefore hold together on a dense open set. The Cayley transform of rational skew-Hermitian matrices gives a dense set of rational unitaries, so a rational quadruple passes all tests. Fix one such quadruple and a common denominator $M_0$ of the entries of all $U_l\hat\sigma_jU_l^\dagger$.

\emph{Shifting Paulis.}
Label basis states by $\mathbb F_D$ with absolute trace $\tr_2$. Up to sign, $Q_{x,z}|u\rangle=i^{\tr_2(xz)}(-1)^{\tr_2(zu)}|u+x\rangle$. For $n\ge7$ choose a six-dimensional binary subspace $\Lambda$ containing two disjoint triples of elements $h$ with $\sum_h1/h=0$ in each triple. The reciprocals of $a,b,a+b$ for distinct nonzero $a,b$ give one triple. A multiple avoiding its nine pairwise ratios gives another, and the six elements span at most six dimensions. Identify $\Lambda$ with the core basis. Fix an origin and a color $l\in\{1,2,3,4\}$ on each coset and install core $l$ in its coordinates. The direct sum defines one encoded qubit. Permute the basis by $\pi(y)=1/y$, with $\pi(0)=0$.

For $x\ne0$, the frame row $f_{x,z}$ collects matrix elements on pairs $\{y,y+h\}$ with $0\ne h\in\Lambda$ and $\pi(y)+\pi(y+h)=x$. For $y\ne0,h$, this gives $(y/h)^2+y/h=1/(xh)$, with one unordered pair exactly when $\tr_2(1/(xh))=0$. The exceptional pair $\{0,h\}$ occurs only for $h=1/x$. In a triple avoiding $1/x$ the three traces sum to zero, so some $h$ contributes exactly one pair. Pull back $Q_{x,z}$ through $\pi$ and add its coset blocks of each color to get $C_l$. At most $2(d-1)$ unordered pairs contribute Gaussian units. The single pair for that $h$ gives a nonzero entry of $\sum_l C_l$, since the difference of the two core indices determines $h$. Thus $D f_{x,z}=v(\mathbf C)$ with $\mathbf C\in\mathcal C$ for every coloring. Finiteness and (1) give fixed $c_1,C_1>0$ with
\begin{equation}
 |f_{x,z}^\top f_{x',z'}|\ge c_1/D^2,\qquad
 \|f_{x,z}\|^2\le C_1/D^2\qquad(x,x'\ne0).
 \label{eq:sm_shift_rows}
\end{equation}

\emph{Diagonal Paulis.}
For $z\ne0$ a coset $y_0+\Lambda$ contributes $v_{U_l}(\operatorname{diag}s)$ to $D f_{0,z}$, where $s_r=(-1)^{\tr_2[z\pi(y_0+r)]}$. For $h\ne0$,
\[
 \left|\sum_y(-1)^{\tr_2[z(\pi(y)+\pi(y+h))]}\right|\le 2\sqrt D+3 .
\]
Indeed, $v=y/(y+h)$ gives $\pi(y)+\pi(y+h)=(v+v^{-1})/h$ for $y\notin\{0,h\}$. With $a=z/h$ and $X=av$, restoring $v=1$ gives the sum of $(-1)^{\tr_2(X+a^2/X)}$ over $X\ne0$, up to three unit terms. This sum plus $D+1$ counts the points of $Y^2+XY=X^3+a^4$, hence the bound follows from Hasse's theorem~\cite{HasseBound}. Averaging $(\sum_r s_r)^2$ over cosets now gives at most $d+d(d-1)(2\sqrt D+3)/D$. A constant profile contributes $d^2$, so its fraction is at most $1/d+(2\sqrt D+3)/D<1/2$ for $D\ge128$.

Color the cosets independently and uniformly. For each $z\ne0$, some nonconstant profile occurs on $N\ge D/(2d\,2^d)$ cosets. Condition on all other colors. By affine independence in (2), at most one vector of the four color counts can make $f_{0,z}=0$. The largest atom of the equal-probability multinomial distribution is $O(N^{-3/2})$ by Stirling's formula. Summing over $D-1$ rows gives failure probability $O(D^{-1/2})$. For sufficiently large $D$ we can fix a coloring with all rows nonzero. Since $D f_{0,z}\in M_0^{-1}\mathbb Z^3$,
\begin{equation}
 \|f_{0,z}\|\ge1/(M_0D)\qquad(z\ne0).
 \label{eq:sm_diag_rows}
\end{equation}

\emph{Graph gap.}
Let $O=\{(x,z):x\ne0\}$, $J=\{(0,z):z\ne0\}$, and let $\mu$ be the mean rate on $O$. Denote the graph energies inside $O$ and between $J$ and $O$ by $E_O$ and $E_{JO}$. Equation~\eqref{eq:sm_shift_rows} gives
\[
 E_O\ge c_1^2|O|D^{-4}\sum_{a\in O}(t_a-\mu)^2,\qquad |O|=D(D-1).
\]
By Parseval, $\sum_{a\in O}f_af_a^\top$ is the Gram matrix of the off-diagonal parts of the three axes. A basis permutation preserves this matrix. Before permutation it is a convex combination of four positive definite core matrices by (3), hence it dominates $\kappa I_3$ for some fixed $\kappa>0$. With $w_{ba}=(f_b^\top f_a)^2$, Eqs.~\eqref{eq:sm_shift_rows} and~\eqref{eq:sm_diag_rows} and the identity $\sum_a f_af_a^\top=I_3$ give
\[
 \sum_{a\in O} w_{ba}\ge\frac{\kappa}{M_0^2D^2}\quad(b\in J),\qquad
 \sum_{b\in J} w_{ba}\le\frac{C_1}{D^2}\quad(a\in O).
\]
Multiply $(t_b-\mu)^2\le2(t_b-t_a)^2+2(t_a-\mu)^2$ by $w_{ba}$ and sum to obtain
\[
 \sum_{b\in J}(t_b-\mu)^2\le
 \frac{2M_0^2}{\kappa}\left[D^2 E_{JO}+C_1\sum_{a\in O}(t_a-\mu)^2\right].
\]
For centered $t$, $\|t\|_2^2\le\sum_a(t_a-\mu)^2$, so the full energy controls $D^{-2}\|t\|_2^2$ with a uniform positive coefficient. In the finitely many smaller dimensions Sec.~\ref{sec:pauli_orbit} gives a connected encoder. Taking the minimum of the large-dimension bound on $D\gamma$ and its finitely many positive values completes the proof.
\section{Upper bound from the relations among the axes}
\label{sec:universal_redundancy}
Let $F=[F_1\ \cdots\ F_k]\in\R^{m\times3k}$ collect the frames of a family with $3k=m+\nu$ axes. If the axes do not span $V$, take a unit $v$ orthogonal to all of them and a unit $u\perp v$: the response $uv^\top$ is centered, annihilates every plane and has zero leakage, so $g=0$. Otherwise the frame operator $S=FF^\top=\sum_jP_j$ is invertible. Let the $\nu$ orthonormal columns of $R$ span $\ker F$, the space of relations, and let $R_j$ be the three rows of $R$ that belong to qubit $j$. Then $F^\top S^{-1}F=I_{3k}-RR^\top$.

The canonical reconstruction from the plane components $P_jv$ is $v=\sum_j S^{-1}P_jv$. Taking adjoints of its summands gives
\[
 X_j=P_jS^{-1},\qquad\sum_jX_j=I_m .
\]
Each $X_j$ has range in plane $j$. Without relations it would project onto that plane along the others, with zero leakage. In general
\[
 X_jF_\ell=F_j(\delta_{j\ell}I_3-R_jR_\ell^\top),
\]
so $X_j$ maps its own plane into itself and leaks from plane $\ell\ne j$ by at most $\|R_jR_\ell^\top\|_F$. Summing over the family and using $\|RR^\top\|_F^2=\nu$, $\sum_j\tr(R_jR_j^\top)=\nu$ and the block size three,
\begin{equation}
 k\sum_je(X_j)^2\le\nu-\sum_j\|R_jR_j^\top\|_F^2
 \le\nu-\frac{\nu^2}{3k}=\frac{\nu m}{3k}.
 \label{eq:sm_numerator}
\end{equation}
The centered responses $Y_j=X_j-(a_j/m)I_m$, $a_j=\tr X_j$, have the same leakage. Since $a_j=\langle P_j,X_j\rangle_F$ and $\|P_j\|_F^2=3$, the Cauchy--Schwarz inequality gives $\|X_j\|_F^2\ge a_j^2/3$, and $\sum_ja_j=m$ gives
\begin{equation}
 \sum_j\|Y_j\|_F^2\ge\Bigl(\frac13-\frac1m\Bigr)\sum_ja_j^2
 \ge\frac{m(m-3)}{3k}.
 \label{eq:sm_denominator}
\end{equation}
Because $g\|Y_j\|_F\le e(Y_j)$ for every $j$, the ratio of Eqs.~\eqref{eq:sm_numerator} and~\eqref{eq:sm_denominator} bounds the gap of every family,
\begin{equation}
 g^2\le\frac{\nu}{k(m-3)} .
 \label{eq:sm_universal}
\end{equation}
At the threshold, $\nu=6$ and $k=(m+6)/3$,
\begin{equation}
 g_*^2(D,\kc)\le\frac{18}{(m+6)(m-3)} .
 \label{eq:sm_minimum_upper}
\end{equation}
In terms of the fractional redundancy $x=\nu/m$, with $k=m(1+x)/3$ and $m-3=D^2-4$, the bound is the upper envelope of Fig.~3,
\begin{equation}
 Dg_*(D,k)\le U_D(x)=\sqrt{\frac{3x}{(1+x)(1-4/D^2)}} .
 \label{eq:sm_finite_precision}
\end{equation}
\section{Lower bound at the threshold}\label{sec:endpoint}
At $k=\kc$ a spanning frame has $m+6$ columns and six independent relations. We start from the seven planes of the two-qubit core in Sec.~\ref{sec:exact} and distribute these relations over the full frame. The construction gives $g\ge c/m$, matching Eq.~\eqref{eq:sm_minimum_upper} in order. Throughout this section $c,C>0$ are independent of $D$.

\subsection{From relations to leakage}
Let $s=\sqrt{\lambda_{\min}(FF^\top)}>0$. Write the six orthonormal relation vectors as columns of $R$, with three rows $R_j$ per plane, and define
\[
 g_{\rm rel}=\min_{\tr B=0,\,\|B\|_F=1}\|(I-RR^\top)BR\|_F,\qquad
 B=\operatorname{diag}(B_1,\ldots,B_k),\quad B_j\in\R^{3\times3}.
\]
Given $T$, set $B_j=F_j^\top TF_j$, $z=\tr B/(3k)$ and $E=TF-FB$. Then $\|E\|_F^2=ke(T)^2$. Since $ER=-FBR$, we have $\|E\|_F\ge s g_{\rm rel}\|B-zI\|_F$. Each block of $E$ is orthogonal to its plane, so
\[
 s^2\|T-zI_m\|_F^2\le\|(T-zI_m)F\|_F^2=\|B-zI\|_F^2+\|E\|_F^2 .
\]
For centered $T$, $\|T-zI_m\|_F\ge\|T\|_F$. Consequently
\begin{equation}
 \sqrt{k}\,g\ge\frac{s^2g_{\rm rel}}{\sqrt{1+s^2g_{\rm rel}^2}} .
 \label{eq:sm_physicalgap}
\end{equation}

\subsection{A perturbation of the two-qubit core}
The core frame $F^c=[F^c_{\rm coord}\ S^c]$ contains four coordinate triples and three rotated triples. Let $R^c$ span its kernel with orthonormal columns and let $K$ be its last nine rows. The other rows equal $-(F^c_{\rm coord})^\top S^cK$. Since $\|S^c\|\le\sqrt3$, we have $I_6/4\preceq K^\top K\preceq I_6$. For seven blocks $B_c$ and a $6\times6$ matrix $\Lambda$,
\begin{equation}
 \min_z\bigl(\|B_c-zI_{21}\|_F^2+\|\Lambda-zI_6\|_F^2\bigr)^{1/2}
 \le C\|B_cR^c-R^c\Lambda\|_F .
 \label{eq:sm_anchor}
\end{equation}
Indeed, equality $B_cR^c=R^c\Lambda$ induces a map on the quotient by $\ker F^c$ preserving all seven planes. This map is scalar by Sec.~\ref{sec:exact}. As the frame is injective on each triple, $B_c$ and $\Lambda$ have the same scalar value. The least singular value on the complement of this scalar pair is positive and stays positive for arrays near $R^c$.

In the full register the other coordinate axes are $Q_p\otimes Q_v$, where $p$ labels a core Pauli and $v\ne0$ labels an environment Pauli. Each noncore triple in Sec.~\ref{sec:exact} has three distinct environment labels. Keep all coordinate triples fixed. Rotate each supplementary triple by a Gaussian Hamiltonian,
\[
 H_\ell=D^{-1}\sum_{p,v}\omega_\ell(p,v)Q_p\otimes Q_v,\qquad
 A_{\ell a}(\theta)=e^{i\theta H_\ell}(A_{\ell a}^c\otimes I_{D/4})e^{-i\theta H_\ell},
 \quad 1\le\ell,a\le3 ,
\]
where the coefficients are independent standard Gaussians. Write $F_\theta=[F_{\rm coord}\ S_\theta]$. When $\max_\ell\|H_\ell\|\le8$ the nine moving columns change by $O(\theta)$ in Frobenius norm. For small fixed $\theta$ all frame singular values therefore lie in $[s_0,2]$ with $s_0>0$ independent of $D$. At zero angle the frame splits into the core and orthonormal coordinate axes, and $\|F_\theta\|^2\le1+3$ at every angle.

Write $\operatorname{orth}(Y)=Y(Y^\top Y)^{-1/2}$. Let $\Pi_\theta$ project onto the kernel of the three rows of $S_\theta$ on the commuting line. An orthonormal basis of $\ker F_\theta$ is
\[
 Z_\theta=\operatorname{orth}(\Pi_\theta\operatorname{orth}(K)),\qquad
 R_\theta=\operatorname{orth}\binom{-F_{\rm coord}^\top S_\theta Z_\theta}{Z_\theta}.
\]
At zero angle its core rows equal $R^c$. The matrices inverted here have fixed size and singular values bounded below. Taylor expansion for $\|H_\ell\|\le8$ gives, uniformly in $D$,
\[
 \|(R_\theta)_{\rm core}-R^c\|_F\le C\theta,\qquad
 \|(R_\theta)_{\rm noncore}-\theta\Xi\|_F\le C\theta^2 .
\]
The row $(p,v)$ of the Gaussian first derivative $\Xi$ is minus the row of coefficients of the nine commutators $i[H_\ell,A_{\ell a}(0)]$ along $Q_p\otimes Q_v$, multiplied by $K$.

\emph{Independent rotations of the noncore triples.}
Use $\|Y\|_2^2=\tr(Y^\dagger Y)/D$ for register operators. For current axes $A_{ja}$ and relation rows $R_j$, put $O_{j\mu}=\sum_a(R_j)_{a\mu}A_{ja}$, $1\le\mu\le6$. A rotation $\dot A_{ja}=i[X_j,A_{ja}]$ of each noncore triple gives $(\dot FR)_\mu=(\mathcal JX)_\mu$, where
\[
 (\mathcal JX)_\mu=\sum_j i[X_j,O_{j\mu}],\quad
 (\mathcal J^*Y)_j=i\sum_\mu[O_{j\mu},Y_\mu],\quad
 (\mathcal NY)_\beta=\sum_{j,\mu}[O_{j\beta},[O_{j\mu},Y_\mu]] .
\]
Here $\mathcal N=\mathcal J\mathcal J^*$ acts on $V^6$. The bound $\|[O,Y]\|_2\le2\|O\|\|Y\|_2$ gives $\|\mathcal J\|\le C\|R_{\rm noncore}\|_F$. At fixed axes $\mathcal J$ is linear in those rows.

\emph{The normal operator.}
Let $\mathcal N_\Xi$ use the rows of $\Xi$ at the coordinate axes. Split $K$ into three blocks $K_\ell$. Let $h_{\ell t}$ be the vector of coefficients of core Pauli $Q_t$ in the three axes $A^c_{\ell a}$ and put $b_{\ell t}=K_\ell^\top h_{\ell t}$. Orthonormality gives $\sum_{\ell,t}b_{\ell t}b_{\ell t}^\top=K^\top K$. The commutator with $Q_t$ contributes $\pm2\omega_\ell(p+t,v)/D$ when $[p,t]=1$. Row $(p,v)$ thus has covariance $\Sigma_p/D^2$, where
\[
 \Sigma_p=4\sum_{\ell,t:[p,t]=1}b_{\ell t}b_{\ell t}^\top,\quad
 \sum_p\Sigma_p=32K^\top K,\quad
 \sum_{p:[p,p']=1}\Sigma_p\succeq16K^\top K\quad(p'\ne0).
\]
The last two bounds count the eight core Paulis anticommuting with $t$ and the at least four anticommuting with both $t$ and $p'$. Inside each triple, distinct environment labels make its three rows independent. Their cross terms vanish in expectation. The expected operator splits into $6\times6$ blocks on labels $w=(p',y)\ne0$ with values
\[
 N_w=\frac4{D^2}\sum_{p,\,v\ne0:\,[(p,v),w]=1}\Sigma_p .
\]
Put $d_{\rm e}=(D/4)^2$. For $y\ne0$ the sum in this expression equals $(d_{\rm e}/2)\sum_p\Sigma_p-\sum_{p:[p,p']=1}\Sigma_p$, giving $N_w\succeq(4-8/d_{\rm e})K^\top K$. For $y=0$ it equals $(d_{\rm e}-1)\sum_{p:[p,p']=1}\Sigma_p$, giving $N_w\succeq4(1-1/d_{\rm e})K^\top K$. Thus $\mathbb E\mathcal N_\Xi\succeq I/2$ for $D\ge8$.

The 48 coefficients at each $v\ne0$ form an independent vector $\omega_v$. Every triple uses three vectors, and each vector appears in at most 16 triples. The resulting dependency graph has degree at most 45, so its triples can be partitioned into 46 classes, each consisting of independent terms. Truncate $\omega_v$ to zero above norm $b_D=\sqrt{48}+\sqrt{8\log D}$. Gaussian concentration bounds the probability of any truncation by $(d_{\rm e}-1)D^{-4}$~\cite{VanHandel2016}. Each truncated vector has mean zero and covariance $a_DI$ with $a_D\to1$. The expected truncated operator is therefore at least $7I/16$ for large $D$. Each centered triple term has norm $O(b_D^2/D^2)$. Matrix Hoeffding in each class with deviation $1/(8\cdot46)$ bounds the total failure probability for the truncated sum by $2(6m)46\exp(-cD^2/b_D^4)$~\cite[Theorem~1.3]{Tropp2012}. Hence $\mathcal N_\Xi\succeq I/4$ with probability tending to one.

For a unit vector $z$, $z^\dagger H_\ell z$ has variance $1/D$. A $1/4$-net of size at most $9^{2D}$ gives $\max_\ell\|H_\ell\|\le8$ with probability $1-o(1)$. On this event $\|\Xi\|_F\le C$, and the Taylor bound and linearity of $\mathcal J$ in the rows give $\|\mathcal N(R_\theta)-\theta^2\mathcal N_\Xi\|\le C\theta^3$. Each noncore coefficient of $S_\theta$ has mean zero. Conjugation by an environment Pauli reversing its sign preserves the law of $H_\ell$ and fixes the initial axes. Duhamel's formula and $\|\mathrm d H_\ell\|_2=\|\mathrm d\omega_\ell\|/D$ give Lipschitz constant $2\theta/D$ for this coefficient. Gaussian concentration and a union bound over the $9(m-15)$ coefficients give a maximum of $8\theta\sqrt{\log D}/D$ with probability $1-o(1)$. Multiplication by the last nine rows of $R_\theta$, of operator norm at most one, bounds the noncore blocks. Thus for sufficiently small fixed $\theta$ and sufficiently large $D$ one can choose a frame satisfying
\begin{equation}
 \mathcal N(R_\theta)\succeq\theta^2I/8,\qquad
 \max_{j\ {\rm noncore}}\|(R_\theta)_j\|_F\le C\theta\sqrt{\log D}/D .
 \label{eq:sm_initialnormal}
\end{equation}

\subsection{Spreading the six relations}
In every noncore block raise singular values below $\alpha/\sqrt m$ to that value, keeping its singular vectors. The resulting increment $\Delta$ vanishes on the core and satisfies $\|\Delta_j\|_F\le\sqrt3\,\alpha/\sqrt m$, $\|\Delta\|_F\le\alpha$. For $0<\alpha<1/4$ set $R(u)=\operatorname{orth}(R_\theta+u\Delta)$, $0\le u\le1$. Before normalization the singular values lie in $[1-\alpha,1+\alpha]$, so the path has length $O(\alpha)$ and
\begin{equation}
 \min_{j\ {\rm noncore}}\sigma_{\min}(R_j(1))\ge\frac{\alpha}{2\sqrt m},
 \qquad
 \max_{j\ {\rm noncore},\,u}\|R_j(u)\|_F\le
 \rho_D=C\left(\frac{\theta\sqrt{\log D}}D+\frac\alpha{\sqrt m}\right).
 \label{eq:sm_roundedblocks}
\end{equation}
The core rows stay within $C(\theta+\alpha)$ of $R^c$.

Starting at $F(0)=F_\theta$, keep the seven core encoders fixed and rotate the noncore axes according to
\[
 Y=-F\dot R,\qquad X=\mathcal J^*\mathcal N^{-1}Y,\qquad \dot A_{ja}=i[X_j,A_{ja}].
\]
Here $\mathcal J$ uses the current axes and the prescribed $R(u)$. Then $\dot FR=Y$, so $FR=0$ throughout the flow. While $\mathcal N\succeq\theta^2I/16$ and $\|F\|\le3$,
\[
 \sum_j\|X_j\|_2^2=\langle Y,\mathcal N^{-1}Y\rangle
 \le C\|\dot R\|_F^2/\theta^2,\qquad
 \|F(u)-F(0)\|_F\le C\alpha/\theta .
\]
Take $\alpha$ small enough that this displacement is below $\min(s_0/4,1/2)$. The frame singular values then stay in $[3s_0/4,5/2]$. The change of $\mathcal J$ due to the rows at fixed axes is $O(\alpha)$. The change due to the axes at fixed rows is at most $C\sqrt D\,\rho_D\|F(u)-F(0)\|_F$, since $\|A-A'\|\le\sqrt D\|A-A'\|_2$. Consequently
\[
 \|\mathcal N(u)-\mathcal N(0)\|
 \le C\alpha+\frac{C\alpha(\theta\sqrt{\log D}+\alpha)}{\theta\sqrt D}<\theta^2/32
\]
for small fixed $\alpha$ and sufficiently large $D$. These estimates keep $\mathcal N\succeq3\theta^2I/32$ and $\|F\|\le5/2$, extending the flow to $u=1$. Full row rank of the final frame then makes $R(1)$ a basis for all its relations.

Finally let $R=R(1)$, $\Lambda=R^\top BR$ and $\zeta=\|BR-R\Lambda\|_F$. By Eq.~\eqref{eq:sm_anchor} there is a scalar $z$ with $\|B_c-zI\|_F+\|\Lambda-zI\|_F\le C\zeta$. For every other block,
\[
 (B_j-zI_3)R_j=(B_jR_j-R_j\Lambda)+R_j(\Lambda-zI_6).
\]
Summing squared norms, using $\sum_j\|R_j\|_F^2=6$ and Eq.~\eqref{eq:sm_roundedblocks}, gives $\|B-zI\|_F^2\le C(m/\alpha^2)\zeta^2$. Hence $g_{\rm rel}\ge c/\sqrt m$. Since $k=(m+6)/3$, Eq.~\eqref{eq:sm_physicalgap} gives $g\ge c/m$. Positivity of the gap at $\kc$ ensures that the planes are distinct by Theorem~1. For the finitely many smaller $D$, the forcing families of Sec.~\ref{sec:exact} have positive gap. In combination with Eq.~\eqref{eq:sm_minimum_upper} this yields
\begin{equation}
 g_*^2(D,\kc)\asymp m^{-2}\asymp\frac{\nu}{\kc m}\qquad(D=2^n\ge4).
 \label{eq:sm_threshold_scaling}
\end{equation}

\section{Lower bound from logarithmic redundancy}
\label{sec:random_completion}
Append $r=k-p=(\nu+3)/3$ independent Haar encodings to the $p=(m-3)/3$ coordinate qubits of Sec.~\ref{sec:exact}. Write $F_j$ and $P_j$ for the added frames and projectors. Let $E(T)=ke(T)^2$ and $E_r(T)=\sum_{j=1}^r\|(I-P_j)TP_j\|_F^2$.

\emph{Coordinate constraints.}
Responses preserving every coordinate plane form a space $\mathcal K$ consisting of $p$ diagonal $3\times3$ blocks and three unrestricted columns on the commuting line. A centered response splits orthogonally as $T=X+Y$ with $X\in\mathcal K_0=\mathcal K\cap I_m^\perp$ and $Y\in\mathcal K^\perp$. The coordinate planes contribute $\|Y\|_F^2$, and
\begin{equation}
 E(T)=\|Y\|_F^2+E_r(T),\qquad \dim\mathcal K_0=6m-10 .
 \label{eq:sm_split}
\end{equation}

\emph{Moments of one Haar plane.}
The squared leakage of a plane is $\|XP\|_F^2-\|PXP\|_F^2$. Let $F$ be an orthonormal frame for the plane. Stack its three rows on coordinate line $i$ above its three rows on the commuting line to obtain $S_i\in\R^{6\times3}$. Write $\Omega_i=mS_iS_i^\top$. Irreducibility of the adjoint action gives $\mathbb EP=(3/m)I_m$, so $\mathbb E\,\Omega_i=3I_6$. For the second term use the orthogonal projection $\mathcal P(X)=PXP$. By Eq.~\eqref{eq:sm_haar_eigenvalues} the largest eigenvalue $\beta$ of $\mathbb E\,\mathcal P$ on centered responses satisfies
\[
 m^2\beta=\frac{6(1-D^{-2})(2-D^{-1})(D-2)}{D-3}<14\qquad(D\ge16),
\]
and $\mathbb E\,\mathcal P(I_m)=(3/m)I_m$. Let $\mathcal P_0$ be the compression of $\mathcal P$ to $\mathcal K_0$. Then $\mathbb E\,\mathcal P_0\preceq(14/m^2)I$ and $\mathbb E\tr\mathcal P_0\le84/m$. A unit vector $v\in\R^6$, embedded in the six selected rows as $y\in\R^m$, has $v^\top\Omega_iv=m\,y^\top Py$ and $(y^\top Py)^2=\langle yy^\top,\mathcal P(yy^\top)\rangle_F$. The scalar part of $yy^\top$ has squared norm $1/m$, hence
\begin{equation}
 \mathbb E(v^\top\Omega_iv)^2\le3+14=17 .
 \label{eq:sm_fourth_moment}
\end{equation}

\emph{Control on $\mathcal K_0$.}
For each plane use one weight $\omega$ and a smaller weight $\omega_i$ on each block. With $M_1=204$ and $M_2=5712$, set
\[
 \omega=\min\Bigl\{1,\frac{M_2}{m\tr\mathcal P_0}\Bigr\},\qquad
 \omega_i=\omega\min\Bigl\{1,\frac{M_1}{\tr\Omega_i}\Bigr\},\qquad
 \widetilde\Omega_i=\omega_i\Omega_i,\qquad
 \widetilde{\mathcal P}=m\,\omega\,\mathcal P_0 .
\]
A zero denominator gives weight one in the corresponding minimum. Then $0\preceq\widetilde\Omega_i\preceq M_1I$, $0\preceq\widetilde{\mathcal P}\preceq M_2I$, and
\[
 \mathbb E\,\widetilde\Omega_i\succeq2I_6,\qquad
 \mathbb E\,\widetilde{\mathcal P}\preceq\frac{14}mI\preceq\frac18I .
\]
Indeed, by Eq.~\eqref{eq:sm_fourth_moment} the truncation of $\Omega_i$ costs at most $\mathbb E[(v^\top\Omega_iv)\tr\Omega_i]/M_1\le6\cdot17/M_1=1/2$ in each unit direction, and the common weight costs at most $[17\Pr\{m\tr\mathcal P_0>M_2\}]^{1/2}\le(17\cdot84/M_2)^{1/2}=1/2$. For $r$ independent planes the matrix Chernoff bounds~\cite{Tropp2012} give, simultaneously for all lines,
\begin{equation}
 \sum_j\widetilde\Omega_{ij}\succeq rI_6,\qquad
 \sum_j\widetilde{\mathcal P}_j\preceq\frac r2I ,
 \label{eq:sm_chernoff}
\end{equation}
except with probability at most $6pe^{-r/(4M_1)}+(6m-10)e^{-r/(4M_2)}$. A row block of $X\in\mathcal K$ on line $i$ involves only the columns of line $i$ and of the commuting line, so its action on $F_j$ is its action on $S_{ij}$, and the rows on the commuting line see only the last three rows of $S_{ij}$. Since $\omega_{ij}\le\omega_j$, the first bound in Eq.~\eqref{eq:sm_chernoff} gives $\sum_j\omega_j\|XF_j\|_F^2\ge(r/m)\|X\|_F^2$. For $X\in\mathcal K_0$, the second bound gives $\sum_j\omega_j\|P_jXP_j\|_F^2\le[r/(2m)]\|X\|_F^2$. Both terms of the leakage now have the same weight $\omega_j$. Since each leakage is nonnegative and $\omega_j\le1$,
\begin{equation}
 E_r(X)\ge\frac{r}{2m}\|X\|_F^2\qquad(X\in\mathcal K_0).
 \label{eq:sm_kernel_gap}
\end{equation}

\emph{Frame bound on $\mathcal K^\perp$.}
For a unit $y\in\R^m$ the overlap $\xi=y^\top a$ with one encoded axis $a$ is a centered function of the encoder with Lipschitz constant $2/\sqrt D$ in the Hilbert--Schmidt metric on $SU(D)$. The logarithmic Sobolev inequality on $SU(D)$ with constant $2/D$~\cite[Theorem~5.16]{Meckes2019} gives $\mathbb Ee^{z\xi}\le e^{4z^2/D^2}$, hence $\Pr\{|\xi|\ge t\}\le2e^{-mt^2/16}$ and $\mathbb Ee^{m\xi^2/32}\le3$. Apply this to the $r$ encodings and a $1/4$-net of at most $9^m$ unit vectors. For each axis the sum of squared overlaps is at most $32[r\log3+m(1+\log9)]/m$ on the net except with probability $e^{-m}$. The supremum over the sphere is at most twice the maximum on the net. A union bound over the three axes gives, except with probability $3e^{-m}$,
\begin{equation}
 \Bigl\|\sum_jP_j\Bigr\|\le b=640\Bigl(1+\frac rm\Bigr),\qquad
 E_r(Y)\le b\|Y\|_F^2 .
 \label{eq:sm_frame_bound}
\end{equation}

\emph{Gap of the completed family.}
The square root of $E_r$ is a seminorm, so Eqs.~\eqref{eq:sm_split}, \eqref{eq:sm_kernel_gap} and~\eqref{eq:sm_frame_bound} give
\[
 \sqrt{\frac r{2m}}\,\|X\|_F\le\sqrt{E_r(T)}+\sqrt b\,\|Y\|_F
 \le\sqrt{(1+b)E(T)} ,
\]
and with $\|Y\|_F^2\le E(T)$,
\begin{equation}
 kg^2\ge\frac{r}{2m(1+b)+r}
       \ge\frac{r}{36^2(m+r)} .
 \label{eq:sm_complete_gap}
\end{equation}
The total failure probability is at most
\begin{equation}
 6pe^{-r/(4M_1)}+(6m-10)e^{-r/(4M_2)}+3e^{-m},
 \label{eq:sm_failure}
\end{equation}
which tends to zero as $D\to\infty$ uniformly in $r\ge C\log m$, for any $C>4M_2$. The planes are distinct almost surely because their common Haar distribution has no atoms. Indeed, the stabilizer of any encoded plane is a proper closed subgroup of the connected group $SU(D)$.

\emph{All redundancies.}
Let $D$ be large and $C'\log D\le\nu\le m$ for a suitable absolute constant $C'$. Then $r/(m+r)\ge\nu/(4m)$ and Eq.~\eqref{eq:sm_complete_gap} gives $g_*^2\ge\nu/(4\cdot36^2km)$. The following construction covers $\nu\ge m$.
\begin{lemma}\label{lem:larger_cardinality}
For all sufficiently large $D$ and every $k\ge k_0=2m/3$ there are $k$ distinct encoded qubits with $g^2\ge1/(6912D^2)$.
\end{lemma}
At $k_0$ we have $\nu=m$ and $r=(m+3)/3$, and Eq.~\eqref{eq:sm_complete_gap} gives a family with
\[
 D^2g_0^2\ge\frac{(m+1)(m+3)}{864m(4m+3)}\ge\frac1{3456}.
\]
Conjugating all its encoders by one unitary conjugates the response by an orthogonal matrix and preserves the gap. For $k=ak_0+k'$ with $0\le k'<k_0$ take $a$ independently rotated copies and $k'$ further Haar encodings. The planes are distinct almost surely, and the copies alone give $g^2\ge(ak_0/k)g_0^2\ge g_0^2/2$. Since $\nu/k<3$, the lemma gives $g_*^2\ge\nu/(27648km)$ for $\nu\ge m$.

For each of the finitely many smaller $D$, Sec.~\ref{sec:exact} supplies a forcing family at $\kc$, and appending distinct planes keeps $kg^2$ from decreasing. Choose one such family at each $\kc\le k\le k_0$. The minimum of $kmg^2/\nu$ over these finitely many cases is positive. Rotating copies of the $k_0$ family covers $k>k_0$ as above. With the upper bound~\eqref{eq:sm_universal} and Eq.~\eqref{eq:sm_threshold_scaling}, this proves the comparison in Theorem~3 at $\nu=6$ and for every $\nu\ge C'\log D$,
\begin{equation}
 g_*^2(D,k)\asymp\frac{\nu}{km},\qquad\text{equivalently}\qquad (Dg_*)^2\asymp\frac{x}{1+x}.
 \label{eq:sm_theorem3}
\end{equation}
\section{An explicit Clifford+\texorpdfstring{$T$}{T} family}
\label{sec:star_construction}
For every $D=2^n\ge4$ we construct $k_0\le4m-2$ distinct encoded qubits, each encoder using at most one $T$ gate, with $k_0g^2\ge1/6$. Here $k_0g^2$ is the minimum of the total squared leakage
\[
 E(X)=\sum_{j=1}^{k_0}\|(I-P_j)XP_j\|_F^2
\]
over centered responses of unit Frobenius norm. Clifford encoders bound the sum of the squared off-diagonal entries of the response, and single-$T$ encoders compare its diagonal entries $t_a=X_{aa}$ along a tree.

\emph{A tree of anticommuting pairs.}
Fix a nonzero label $h$ and let $S_1=\{b:[h,b]=1\}$ and $S_0=h^\perp\setminus\{0,h\}$. Join $a\in S_0$ to $b\in S_1$ when they anticommute. Each $a$ has degree $D^2/4$ and each $b$ has degree $D^2/4-1$, so Hall's condition holds and every $a$ can be matched to a distinct $b$. Joining every $b$ to $h$ gives a spanning tree on the $m$ labels whose edges are anticommuting pairs and whose branches have length at most two. For $\sum_at_a=0$,
\begin{equation}
 \sum_at_a^2\le\sum_a(t_a-t_h)^2
 \le3\sum_{\{a,b\}\in\rm tree}(t_a-t_b)^2 ,
 \label{eq:sm_tree}
\end{equation}
where a branch of length two uses $(t_a-t_h)^2\le2(t_a-t_b)^2+2(t_b-t_h)^2$.

\emph{Two tilted planes per edge.}
For a tree edge $\{a,b\}$ choose a label $c\ne a+b$ that anticommutes with $a$ and $b$; there are $D^2/4-1$ choices. Let $C=Q_c$ and $A_\pm=(Q_a\pm Q_b)/\sqrt2$. The triples $(A_\pm,C,iA_\pm C)$ are obtained from the coordinate qubit generated by $Q_a$ and $C$ by the rotations $\exp(\mp i\pi H/8)$ with $H=iQ_bQ_a$, which commutes with $C$ and turns $Q_a$ into $A_\pm$. A Clifford maps $H$ to $Z$ on one wire, so each rotation is one $T$ gate up to Cliffords. The projector of a tilted plane has diagonal entries $1$ at $c$ and $1/2$ at $a$, $b$, $a+c$, $b+c$, and off-diagonal entries only on the pairs $\{a,b\}$ and $\{a+c,b+c\}$.

The unit vector $B_\pm=(Q_b\mp Q_a)/\sqrt2$ is orthogonal to the plane of $A_\pm$, so the leakage of that plane is at least $\langle B_\pm,XA_\pm\rangle^2$, and
\[
 \langle B_\pm,XA_\pm\rangle
 =\tfrac12(X_{ba}-X_{ab})\pm\tfrac12(t_b-t_a).
\]
The two squares sum to at least $(t_b-t_a)^2/2$. By Eq.~\eqref{eq:sm_tree} the tilted planes contribute at least $\sum_at_a^2/6$ to $E(X)$ when $\tr X=0$.

Two tilted planes from different edges coincide only if they use the same $c$ and the edges are translates of each other by $c$. Call $c$ an exit of the edge $e$ when $e+c$ is not a tree edge, and otherwise link $e$ to $e+c$. This defines a simple graph on the tree edges: $c$ anticommutes with the translated endpoints as well, $e+c\ne e$ because $c\ne a+b$, and two shifts with the same neighbor differ by $a+b$, so only one of them anticommutes with $a$. An edge without an exit has $D^2/4-1\ge3$ neighbors. In a component that contains an edge with an exit, let that edge take its exit and let every other edge choose its parent in a spanning tree rooted there. A component without exits contains a cycle. Orient the cycle and let every other edge choose its parent in a forest rooted on the cycle. No two edges choose each other, so the $2(m-1)$ tilted planes are distinct.

\emph{Coordinate planes.}
Each nonzero label $a$ lies on $D^2/4$ qubit lines $\{a,b,a+b\}$. Choose two of them for every $a$. They meet only in $a$, so every $b\ne a$ lies outside one chosen line through $a$, and the leakage of that coordinate plane contains $X_{ba}^2$. The union of the chosen lines has at most $2m$ Clifford planes. For a coordinate plane, the leakage is a sum of squared entries, so the union contributes at least $\sum_{a\ne b}X_{ba}^2$. Altogether
\begin{equation}
 k_0\le2m+2(m-1),\qquad
 E(X)\ge\sum_{a\ne b}X_{ba}^2+\frac16\sum_at_a^2
       \ge\frac16\|X\|_F^2\qquad(\tr X=0),
 \label{eq:sm_star_gap}
\end{equation}
and therefore $1/g\le\sqrt{6k_0}<5D$.
\section{A perfectly transmitted bit at small RMS leakage}
\label{sec:hidden_bit}
Fix a finite family $\mathcal F$ of encoded qubits. For distinct inputs, define
\[
 r_\Phi(\rho,\sigma)=\frac{\|\Phi(\rho)-\Phi(\sigma)\|_1}{\|\rho-\sigma\|_1}.
\]
Let $K_{\rm ret}(\mathcal F)$ be the least constant such that $\bigl|r_\Phi-|\lambda|\bigr|\le K_{\rm ret}e(T)$ for every channel and state pair. For unital channels, define $K_\diamond(\mathcal F)$ by the analogous bound on $\inf_{-1/m\le z\le1}\|\Phi-\mathcal D_z\|_\diamond$, using the full diamond norm.

\emph{One-bit witness.}
Every finite family admits a unital entanglement-breaking channel with
\begin{equation}
 \lambda=0,\qquad e(T)^2\le\frac{6}{Dm},\qquad
 \max_{\rho\ne\sigma}r_\Phi=1,\qquad
 \min_{\rho\ne\sigma}r_\Phi=0 .
 \label{eq:sm_hidden_bit}
\end{equation}
To construct it, let $\bar w_a$ be the Pauli projection weight averaged over the family. Since $\sum_a\bar w_a=3$, some $Q=Q_a$ has $\bar w_a\le3/m$. Choose orthogonal levels $|0\rangle,|1\rangle$ in its positive eigenspace and set $S=|0\rangle\langle0|-|1\rangle\langle1|$. Measure $(I\pm S)/2$ and prepare $(I\pm Q)/D$ conditioned on the outcome. The channel is
\begin{equation}
 \Psi(Z)=\frac{\tr(Z)I+\tr(SZ)Q}{D}.
 \label{eq:sm_bit_channel}
\end{equation}
The two levels produce orthogonal outputs and transmit one classical bit without error. The superpositions $(|0\rangle\pm|1\rangle)/\sqrt2$ both produce $I/D$.

In the Heisenberg picture, $\Psi^\dagger(Q)=S$ and every other nonidentity Pauli maps to zero. The response is $T=s\mathbf e_a^\top$, with $s_b=\tr(Q_b S)/D$. Thus $\tr T=s_a=0$ and $\|s\|_2^2=2/D$. Plane $j$ contributes $\|(I-P_j)s\|_2^2(P_j)_{aa}$ to the squared leakage before averaging, which gives Eq.~\eqref{eq:sm_hidden_bit}. Since $\lambda=0$ and $R=\|s\|_2=\sqrt{2/D}$, the level pair has $r_\Psi=1=\sqrt{D/2}\,R$. It attains equality in $\bigl|r_\Phi-|\lambda|\bigr|\le\sqrt{D/2}\,R$ of Sec.~\ref{sec:setup}, so the factor $\sqrt{D/2}$ is sharp.

The distance to the nearest depolarizing channel is exactly one. Since $\|S/2\|_1=1$ and $\tr[Q(\Psi-\mathcal D_z)(S/2)]=1$ for every $z$, the full diamond distance is at least one throughout the family. Complete depolarization attains the bound because $\Psi-\mathcal D_0$ maps $Z$ to $\tr(SZ)Q/D$ and has diamond norm $\|S\|_\infty\|Q/D\|_1=1$. Indeed the adjoint of the functional $Z\mapsto\tr(SZ)$ tensors any auxiliary test observable with $S$.

Conversely, the difference of an erased pair belongs to $\ker T^\top$, so $R\ge|\lambda|$. A perfectly transmitted pair gives $1-|\lambda|\le\sqrt{D/2}\,R$ by the trace norm bound of Sec.~\ref{sec:setup}. Adding these inequalities yields
\[
 e(T)\ge\frac{g}{1+\sqrt{D/2}} .
\]
For families with $g=\Theta(1/D)$, the construction therefore attains the smallest possible leakage order $D^{-3/2}$ for this combination of perfect transmission and complete erasure.

\emph{Weak noise.}
For $\Phi_p=(1-p)\operatorname{id}+p\Psi$, $0<p<1$, the mean rate is $\lambda=1-p$ and $e(T_{\Phi_p})=p\,e(T_\Psi)$. The difference of the two level states becomes $(1-p)S+2pQ/D$. When $1-p\ge2p/D$, its two eigenvalues on the codeword subspace have opposite signs or one is zero. Each of the other $D-2$ eigenvalues has magnitude $2p/D$. The two retentions are therefore
\[
 r_{\rm levels}=1-p+p(1-2/D),\qquad r_{\rm superpositions}=1-p .
\]
The excess retention of the level states is $p(1-2/D)$. It scales with leakage, so the amplification factor persists arbitrarily close to the identity.

\emph{Sharp dimension dependence.}
Every family with $g>0$ satisfies
\begin{align}
 \sqrt{\frac{Dm}{6}}&\le K_{\rm ret}(\mathcal F)
 \le\frac{\sqrt{D/2}}{g},
 \label{eq:sm_bit_retention}\\
 \sqrt{\frac{(D-1)m}{3}}&\le K_\diamond(\mathcal F)
 \le\frac{\sqrt D}{g}.
 \label{eq:sm_bit_diamond}
\end{align}
The first lower bound follows from Eq.~\eqref{eq:sm_hidden_bit}, and the first upper bound is Eq.~\eqref{eq:sm_retention_conversion}.

For the diamond lower bound, replace $S$ in Eq.~\eqref{eq:sm_bit_channel} by
\[
 H=\frac{D|w\rangle\langle w|-I}{D-1},
\]
where $\langle w|Q|w\rangle=0$. The eigenvalues of $H$ are $1$ and $-1/(D-1)$, so $(I\pm H)/2$ define a measurement. Here $\tr(HQ)=0$ and $\tr(H^2)/D=1/(D-1)$, giving $e^2\le3/[m(D-1)]$. On input $|w\rangle$, the output has unit $Q$ expectation, while every $\mathcal D_z$ gives zero. This proves the lower bound.

For the upper bound, $\mathcal D_\lambda$ is physical because the entanglement fidelity $(1+m\lambda)/D^2$ lies in $[0,1]$. Write $\mathcal E=\Phi-\mathcal D_\lambda$. For unital $\Phi$, reshuffling the centered Liouville block gives $\|J(\mathcal E)\|_F=R$. In the Choi variational formula for the diamond norm~\cite{Watrous2018,BitWatrous}, the left and right factors multiplying $J(\mathcal E)$ are $\sqrt\rho\otimes I$ and $\sqrt\sigma\otimes I$ for input density operators $\rho,\sigma$. Their Frobenius and operator norms are respectively $\sqrt D$ and at most one. H\"older's inequality gives $\|\mathcal E\|_\diamond\le\sqrt D\,R$, and Eq.~\eqref{eq:sm_channel_stability} completes the bound.

At fixed positive fractional redundancy, Theorem~3 shows that the optimal constants in both tasks scale as $\Theta(D^{3/2})$.
\space
\section{Finite-shot cost}
\label{sec:shot_cost}
Each call uses the same memoryless channel, calibrated controls and one decoded logical Pauli measurement. We test $R=0$ against $R\ge\epsilon$ with each error probability at most $\delta$, where $0<\epsilon\le1$ and $0<\delta<1/2$. For a family with gap $g\ge g_0>0$, the test uses
\begin{equation}
 \widetilde O\!\left(\frac{D}{g_0^2\epsilon^2}\right)
 \label{eq:sm_shot_cost}
\end{equation}
calls, with logarithmic dependence on $D$, $1/g_0$, $1/\epsilon$ and $1/\delta$.

\emph{One shot.}
Choose an encoder $U_j$ uniformly from the family and an axis uniformly from its three logical axes. Draw a logical pure state $z$ and two spectator states $\psi_\pm$ independently from projective 2-designs in their respective spaces~\cite{Dankert2009}. Together these choices specify a setting. On each repetition draw a fresh fair sign $s$, prepare $U_j(z\otimes\psi_s)U_j^\dagger$, apply the channel, decode, and multiply the Pauli outcome by $s$. The inputs have the same logical marginal. The mean $\mu$ at this setting is half the difference between the two mean readouts and vanishes under the null.

Expand the adjoint image of the measured Pauli in the encoded basis $\sigma_\alpha\otimes Q_b$, where $\sigma_0=I$. Its coefficients $c_{\alpha b}$ for $b\ne0$ form one leakage column. With $z_\alpha=\tr(z\sigma_\alpha)$ and spectator dimension $d=D/2$, the identity $\mathbb E\,\psi^{\otimes2}=(I+\mathsf{Swap})/[d(d+1)]$ gives
\[
 \mathbb E_{\psi_\pm}\mu^2
 =\frac1{2(d+1)}\sum_{b\ne0}\left(\sum_{\alpha=0}^3 c_{\alpha b}z_\alpha\right)^2 .
\]
The logical Bloch vector has zero mean and covariance $I_3/3$. Averaging over $z$, the three axes and the family yields
\begin{equation}
 \mathbb E\mu^2\ge\frac{e(T)^2}{18(d+1)}
 \ge\frac{g_0^2R^2}{18(d+1)} .
 \label{eq:sm_shot_moment}
\end{equation}
Thus the alternative has second-moment signal at least $q_0=g_0^2\epsilon^2/[18(d+1)]$.

\emph{Multiscale repetition.}
Use $L=\lceil\log_2(2/q_0)\rceil$ levels with $t_\ell=2^{-\ell}$, $1\le\ell\le L$, as in multiscale tests of Pauli measurement outcomes~\cite{ReadoutYu}. At level $\ell$, draw $M_\ell$ independent settings and repeat each $N_\ell$ times, with
\[
 M_\ell=\left\lceil \frac{4Lt_\ell}{q_0}\log\frac2\delta\right\rceil,\qquad
 M=\sum_{\ell=1}^LM_\ell,\qquad
 N_\ell=\left\lceil\frac{32}{t_\ell}\log\frac{4M}{\delta}\right\rceil .
\]
Reject if any sample mean at level $\ell$ exceeds $\sqrt{t_\ell}/2$ in absolute value. Hoeffding's inequality bounds every estimation error by $\sqrt{t_\ell}/4$ with joint probability at least $1-\delta/2$. Under the null all means vanish. Under the alternative the settings with $\mu^2<t_L$ contribute at most $q_0/2$ to $\mathbb E\mu^2$. Some dyadic bin $[t_\ell,2t_\ell]$ therefore carries at least $q_0/(2L)$ of the second moment and has probability at least $q_0/(4Lt_\ell)$. The probability of missing this bin in $M_\ell$ draws is at most $\delta/2$. Hitting the bin and estimating its mean to the stated accuracy suffices for rejection.

The total number of calls is $\sum_\ell M_\ell N_\ell=O(q_0^{-1}L^2\log(2/\delta)\log(4M/\delta))$, giving Eq.~\eqref{eq:sm_shot_cost}. At each fixed positive fractional redundancy, Theorem~3 gives families with $g_0\gtrsim1/D$, for which the cost is $\widetilde O(D^3/\epsilon^2)$. The Clifford+$T$ family in Sec.~\ref{sec:star_construction} achieves this cost explicitly.
\section{How much does a fixed logical readout cost?}
\label{sec:minimax_readout}
Each round sends a fresh $D$-dimensional register through the same memoryless channel and returns one classical bit. The input is a state of the register alone. The experimenter retains only classical records and may use private randomness and feedback to choose the next input and measurement. A fixed readout uses effects in one encoded algebra, chosen before the first round. A switchable readout may use a different algebra each round. The switchable lower bounds hold for arbitrary binary effects on the register.

\emph{Minimax statement.}
Fix an error probability $0<\delta\le1/3$ independent of $D$ and $\epsilon$. For $0<\epsilon\le1/4$, the smallest worst-case call budgets for testing $R=0$ against $R\ge\epsilon$ with each error at most $\delta$ satisfy
\begin{equation}
\begin{aligned}
 N_*(\text{general, switchable})&=\widetilde\Theta(D^3/\epsilon^2),\\
 N_*(\text{Pauli, fixed})&=\widetilde\Theta(D^3/\epsilon^2),\\
 N_*(\text{Pauli, switchable})&=\widetilde\Theta(D^2/\epsilon^2).
 \end{aligned}
 \label{eq:sm_minimax_readout}
\end{equation}
The tilde accounts for logarithmic factors in the upper bounds. Each test compares preparations with the same logical marginal and reads one decoded logical Pauli. The Pauli results also hold for alternatives satisfying $\inf_z\|\Phi-\mathcal D_z\|_\diamond\ge\epsilon$.

\emph{Hidden input and output axes.}
For nonidentity Paulis $Q_a,Q_b$, define
\begin{equation}
 \Phi_{ab}(\rho)=\frac{I+\eta\,\tr(\rho Q_b)Q_a}{D},
 \qquad0<\eta\le1/2 .
 \label{eq:sm_minimax_alts}
\end{equation}
This channel measures $Q_b$ and prepares $(I\pm\eta Q_a)/D$ with the measured sign. For $a\ne b$, the response has one off-diagonal entry, $R=\eta$, and the null is $\mathcal D_0$. For $a=b$, this is a Pauli channel with one nonzero nonidentity rate. The matched null is $\mathcal D_{\eta/m}$. By Lemma~\ref{lem:nearest_pauli},
\[
 R=\eta\sqrt{1-1/m},\qquad
 \inf_z\|\Phi_{aa}-\mathcal D_z\|_\diamond=\eta(m-1)/m .
\]
Take $\eta=\epsilon$ for general alternatives and $\eta=2\epsilon$ for Pauli alternatives.

Condition on the tester's private seed and the classical record up to the current round. For its next input $\rho$ and output effect $E$, put
\[
 p=\frac{\tr E}{D},\qquad r_b=\tr(\rho Q_b),\qquad
 z_a=\frac{\tr(EQ_a)}D .
\]
When $p=0$ or $1$, the outcome is deterministic under every channel and contributes zero information. Otherwise let $s_a=z_a/\sqrt{p(1-p)}$. Parseval and $E^2\preceq E$ give
\begin{equation}
 \sum_br_b^2=D\tr(\rho^2)-1\le D-1,
 \qquad \sum_as_a^2\le1 .
 \label{eq:sm_minimax_parseval}
\end{equation}
Under $\Phi_{ab}$, the outcome probability is $p'=p+\eta r_bz_a$. Since $|z_a|\le\min(p,1-p)$ and $|r_b|\le1$, we have $p'(1-p')\ge p(1-p)/4$. Under the null, $p_0=p$ for $a\ne b$ and $p_0=p+(\eta/m)\sum_b r_bz_b$ for $a=b$. The binary relative entropy bound $\mathrm{kl}(p_0\Vert p')\le(p_0-p')^2/[p'(1-p')]$ gives respective one-round bounds $4\eta^2r_b^2s_a^2$ and $4\eta^2(r_as_a-\bar h)^2$, with $\bar h=m^{-1}\sum_b r_bs_b$.

Let $\mathbb P_0,\mathbb P_{ab}$ denote the laws of the full record, including the seed. The chain rule for $D_{\rm KL}(\mathbb P_0\Vert\mathbb P_{ab})$ takes expectations over histories under the common null. We can thus average over the hidden axes with each history held fixed. Data processing to the final decision gives $D_{\rm KL}(\mathbb P_0\Vert\mathbb P_{ab})\ge\mathrm{kl}(1-\delta\Vert\delta)$ for every alternative.

\emph{Switchable readouts.}
Averaging over uniform $a\ne b$ gives mean information $O(\eta^2(D-1)/[m(m-1)])$ per round. A successful general-channel test therefore requires
\[
 N\ge\mathrm{kl}(1-\delta\Vert\delta)\,
 \frac{m(m-1)}{4\eta^2(D-1)} .
\]
For uniform Pauli alternatives $a=b$, centering by $\bar h$ and Eq.~\eqref{eq:sm_minimax_parseval} give mean information at most $4\eta^2/m$, hence
\[
 N\ge\mathrm{kl}(1-\delta\Vert\delta)\frac{m}{4\eta^2}.
\]
These are the switchable lower bounds.

\emph{One fixed logical qubit.}
Fix the private seed $\omega$. For its encoder, put $w_a=P_{aa}$, so $\sum_aw_a=3$. Every logical effect has $Ps=s$ and $\|s\|_2\le1$, hence $|s_a|\le\sqrt{w_a}$. Write $n_a=\sum_{t=1}^N\mathbb E_0^\omega r_{t,a}^2$ for the input weight accumulated under the common null. Then $\sum_an_a\le N(D-1)$ and $|\bar h_t|\le\sqrt{D-1}/m$ in each round. Pinsker's inequality, the chain rule and the triangle inequality in $L^2$ give
\[
 \mathrm{TV}(\mathbb P_0^\omega,\mathbb P_{aa}^\omega)
 \le \sqrt2\,\eta\left(\sqrt{w_an_a}+\frac{\sqrt{N(D-1)}}m\right).
\]
The hidden axis is uniform and independent of the seed. Since $\sum_a\sqrt{w_an_a}\le\sqrt{3N(D-1)}$, averaging over axes and then over seeds gives
\begin{equation}
 1-2\delta\le\frac1m\sum_a\operatorname{TV}(\mathbb P_0,\mathbb P_{aa})
 \le\sqrt2\,\eta(\sqrt3+1)\frac{\sqrt{N(D-1)}}m .
 \label{eq:sm_minimax_fixed}
\end{equation}
The left side follows from the acceptance probabilities under the two hypotheses. With $\eta=2\epsilon$, a fixed encoder therefore requires $\Omega(D^3/\epsilon^2)$ calls. For a bank of $k$ encoders chosen in advance, take $w_a=\max_{j\le k}(P_j)_{aa}$. Their sum is at most $3k$. Combining this bound with the switchable lower bound gives $\Omega(\max\{D^2,D^3/k\}/\epsilon^2)$.

\emph{Matching autonomy tests.}
For general channels, use the Clifford+$T$ family of Sec.~\ref{sec:star_construction}. Its bounds $k_0g^2\ge1/6$ and $k_0\le4m-2$ give $\mathbb E\mu^2\gtrsim R^2/D^3$ by Eq.~\eqref{eq:sm_shot_moment}. For Pauli noise, the fixed encoder of Sec.~\ref{sec:flat_all} has $e(T)\ge\gamma R$ with $\gamma\gtrsim1/D$ and gives the same second moment bound. Multiscale repetition yields the two $\widetilde O(D^3/\epsilon^2)$ upper bounds.

For switchable Pauli readout, sample a uniform edge $\{a,b\}$ of the anticommutation tree from Sec.~\ref{sec:star_construction}. Set
\[
 L=\frac{Q_a+Q_b}{\sqrt2},\qquad K=\frac{Q_a-Q_b}{\sqrt2}.
\]
The tilted plane has axes $L,C,iLC$ as in Sec.~\ref{sec:star_construction}. Here $C=Q_c$ anticommutes with both $Q_a$ and $Q_b$ and $c\ne a+b$, so $K$ is orthogonal to all three axes. Prepare $(I\pm K)/D$ with a fair sign and measure $L$. These inputs have identical logical marginals, and multiplying the outcome by the sign gives mean
\[
 \mu=\frac{t_a-t_b}{2},\qquad
 \mathbb E_{\rm edge}\mu^2\ge\frac{R^2}{12(m-1)} .
\]
Here the mean square follows from Eq.~\eqref{eq:sm_tree}. Each encoder uses at most one $T$ gate. Multiscale repetition gives the $\widetilde O(D^2/\epsilon^2)$ upper bound. Since $\inf_z\|\Phi-\mathcal D_z\|_\diamond\le(\sqrt m/D)R\le R$ for Pauli channels, the same tests apply to alternatives separated in diamond norm.

\emph{Known mean attenuation near one.}
Suppose $\lambda_\star=1-h$ is known, with fixed $0<h\le1/2$ and $\epsilon\le h/4$. Replace each null and its alternatives by $(1-\alpha)\operatorname{id}+\alpha\Phi$. For $a\ne b$ take $\alpha=h$ and $\eta=\epsilon/\alpha$. For $a=b$ take $\alpha=h+2\epsilon/m$ and $\eta=2\epsilon/\alpha$. Both choices give the null $\mathcal D_{\lambda_\star}$ and the same mean rate for every alternative. The centered response scales by $\alpha$. The Pauli alternatives have diamond separation $\alpha\eta(m-1)/m\ge\epsilon$.

For a fixed input and effect, mixing multiplies the difference in outcome probabilities by $\alpha$. The alternative probability $p'_\alpha$ satisfies $p'_\alpha(1-p'_\alpha)\ge\alpha p'(1-p')$ by concavity. Thus $\alpha\eta^2$ replaces $\eta^2$ in the one-round relative entropy bounds. The same weight budgets give lower bounds $\Omega(hD^3/\epsilon^2)$ for general channels and for Pauli noise with a fixed encoder, and $\Omega(hD^2/\epsilon^2)$ for switchable Pauli readout. Together with the autonomy tests, these bounds give the powers of $D$ and $1/\epsilon$ in Eq.~\eqref{eq:sm_minimax_readout} for every fixed $h$ in this range.
\section{Data and certificates for Figs.~1--3}
\label{sec:numerics}

\emph{Compatible responses in Fig.~1(c).}
At $D=4$, an encoded plane has three logical axes and twelve orthogonal directions. Let $F_j$ and $F_j^\perp$ contain orthonormal bases for these two spaces. Stacking the equations $(F_j^\perp)^\top T F_j=0$ for the first $k$ encoders gives a $36k\times225$ matrix. In the Haar sample of Fig.~1(c), its nullities for $k=1,\ldots,7$ were $189,153,117,81,45,9,1$, using a singular value cutoff of $10^{-9}$ times the largest singular value. Thus the sixth encoder still leaves nine response parameters free. The seventh leaves only the common rate. For the explicit family in Fig.~1(d), \path{data/phase_flip/results.json} gives the modular ranks and the exact readout contrast derived in Sec.~\ref{sec:phase_flip_example}.

\emph{One encoder in Fig.~2.}
The file \path{figure_source/fig2_pauli_edges.csv} lists the 105 edge weights in panel (c). For panel (d), each Haar encoder is the unitary factor in the QR decomposition of a complex Gaussian matrix, with the diagonal of $R$ chosen positive. Table~\ref{tab:haar_seed_medians} compares the gap $\gamma$ with the bound $\gamma_w$ from Eq.~\eqref{eq:sm_min_degree}. The medians of the ratios use paired values from each draw. The individual draws are in \path{data/pauli_seed}.

\begin{table}[htb]
\caption{Haar medians in Fig.~2(d), from 512 encoders at each dimension. The interval for the population median of $\gamma$ uses order statistics 234 and 279 (coverage $95.34\%$). Interval endpoints are rounded outwards.}
\label{tab:haar_seed_medians}
\begin{ruledtabular}\begin{tabular}{rcccc}
$D$ & $\gamma$ & Interval for the median & $D\gamma$ & $\gamma/\gamma_w$ \\ \hline
4  & 0.1030   & $[0.0945,0.1098]$ & 0.412 & 0.9868 \\
8  & 0.04460  & $[0.0426,0.0464]$ & 0.357 & 0.9988 \\
16 & 0.01393  & $[0.0133,0.0146]$ & 0.223 & 0.99988 \\
32 & 0.004215 & $[0.0040,0.0044]$ & 0.135 & 0.99999
\end{tabular}\end{ruledtabular}
\end{table}

\emph{Bounds on the optimal gap in Fig.~3.}
Each interval in Table~\ref{tab:precision-cuts} combines a lower bound from one explicit family with the upper bound $U_D(x)$ from Eq.~\eqref{eq:sm_finite_precision}, valid for every family of that size. The rows marked for plotting in \path{figure_source/fig3_response_precision.csv} identify the eight families. Their integer encoder matrices and spectral cuts are in \path{data/response_precision} and \path{data/extensions}.

\begin{table}[htb]
\caption{Bounds on $Dg_*(D,k)$ in Fig.~3, rounded outwards. The threshold $\kc$ is 23 at $D=8$ and 87 at $D=16$.}
\label{tab:precision-cuts}
\begin{ruledtabular}\begin{tabular}{lcc}
$k$ & $D=8$ & $D=16$ \\ \hline
$\kc$ & $[0.2492,0.5276]$ & $[0.1152,0.2647]$ \\
$\kc+3$ & $[0.4179,0.7845]$ & $[0.1523,0.4115]$ \\
$\kc+9$ & $[0.6900,1.0489]$ & $[0.3179,0.5910]$ \\
$2\kc$ & $[0.9867,1.3188]$ & $[0.9980,1.2486]$
\end{tabular}\end{ruledtabular}
\end{table}

For each family, the Cayley transform $(2^{20}I-K)^{-1}(2^{20}I+K)$ fixes each encoder exactly from its integer matrix $K$. Set $S=\sum_jP_j$, $W=[F_1\otimes F_1\ \cdots\ F_k\otimes F_k]$, and $\iota=\operatorname{vec}I_m$. The leakage operator multiplied by $k$ is $S\otimes I_m-WW^\top$. Adding $\iota\iota^\top$ lifts the scalar eigenvalue to $m$ and leaves the centered subspace invariant with its original spectrum. For a proposed lower bound $g_0$, put $\xi=kg_0^2<m$ and $A=(S-\xi I_m)\otimes I_m+\iota\iota^\top$. A Schur complement gives $g>g_0$ if
\[
  S-\xi I_m\succ0,\qquad I_{9k}-W^\top A^{-1}W\succ0 .
\]
For each of the two matrices $M$, the certificate uses an approximate inverse Cholesky factor $Y$ and bounds the maximum absolute row sum of $I-YMY^\top$ strictly below one. 
\bibliographystyle{apsrev4-2}
\makeatletter
\immediate\write\@auxout{\string\citation{apsrev42Control}}
\makeatother
\bibliography{bibliography-control,references}